# Mass Action
# and
# Conservation of Current

Bob Eisenberg

Department of Molecular Biophysics and Physiology

Rush University Medical Center

Chicago IL 60612

USA

*August 11, 2015*

**File name: "ArXiv Aug 11-2 2015 Mass Action and Conservation of Current.docx"**

# Executive Summary

The law of mass action is used widely, nearly universally, in chemistry to describe chemical reactions. The law of mass action does not automatically conserve current, as is clear from the mathematics of a simple case, chosen to illustrate the issues involved. If current is not conserved in a theory, charges accumulate that cannot accumulate in the real world. In the real world, tiny charge accumulation—much less than one per cent—produces forces that change predictions of the theory a great deal. Indeed, in the real world tiny charge accumulation produces forces large enough to destroy biological membranes and thus living systems, forces large enough to ionize atoms, to create a plasma of electrons (like sparks or lightning) and thus make experiments impossible in normal laboratory settings.

The mathematics in this paper shows that the law of mass action violates conservation of current when current flows if the rate constants depend only on the potential (chemical and electrical) in one location and its immediate vicinity. The same difficulty arises when rate models of the Markov type deal with the movement of charge. The implication is that such models cannot deal with current through an open channel, or with the gating properties of channels if the gating mechanism is charged, or with the gating current produced by that mechanism for example.

The essential issue is that rate constants are LOCAL functions of potential (at one place or in a small region) so they cannot know about current flow far away or at boundary conditions. If current is interrupted far away, local chemical reactions obviously change, but rate constant models show no change. Consider for example what happens in a battery when current flow is interrupted far from the battery.

These problems can be fixed by computing the potential (chemical and electrical $\phi(x)$) GLOBALLY and using the neighborhood values of potential that are the result of the GLOBAL calculation. Rate constants in this case have simple physical interpretation and simple expressions involving only one integration, explained near eq. (19).

$$J_k = \underbrace{C_k\left(L\right)\left(\frac{D_k}{l}\right)\text{Prob}\{R|L\}}_{\text{Unidirectional Efflux}} - \underbrace{C_k\left(R\right)\left(\frac{D_k}{l}\right)\text{Prob}\{L|R\}}_{\text{Unidirectional Infflux}}$$

where $C_k(L)$ is Source Concentration, $\frac{D_k}{l}$ is Diffusion Velocity, $\text{Prob}\{R|L\}$ is Conditional Probability, and $l$ is Channel Length.

$$\text{Prob}\{R_{ight}|L_{eft}\} = \frac{\exp(z_k F V_{trans}/RT\ )}{\frac{1}{l}\int_0^l \exp(z_k F\phi(x)/RT)dx};$$

$$\text{Prob}\{L_{eft}|R_{ight}\} = \frac{1}{\frac{1}{l}\int_0^l \exp(z_k F\phi(x)/RT)dx}.$$

Of course, the rate model and the global potentials interact and must be solved together so they are consistent. Variational methods guarantee such consistency. Energetic variational methods are needed if the systems are dissipative. (Ions in water and channels are dissipative systems because they are condensed phases with little empty space. When ions move, they collide with atoms and dissipate energy in the form of heat.)

**Abstract**

The law of mass action does not force a series of chemical reactions to have the same current flow everywhere. Interruption of far-away current does not stop current everywhere in a series of chemical reactions (analyzed with the law of mass action), and so does not obey Maxwell's equations. An additional constraint and equation is needed to enforce global continuity of current. The additional constraint is introduced in this paper in the special case that the chemical reaction describes spatial movement through narrow channels. In that case, a fully consistent treatment is possible using different models of charge movement. The general case must be dealt with by variational methods that enforce consistency of all the physical laws involved.

The law of mass action does not automatically conserve current, as is clear from the mathematics of a simple case, chosen to illustrate the issues involved. Consider the chemical reaction $X \rightleftharpoons Y \rightleftharpoons Z$: the currents in the two chemical reactions are not equal. $I_{XY} \neq I_{YZ}$. The difference in current is $I_{XY} - I_{YZ} = z_X k_{xy} F[X] - z_Y k_{yx} F[Y] - z_Y k_{yz} F[Y] + z_Z k_{zy} F[Z]$. The difference can be zero only for special circumstances.

Violations of current continuity arise away from equilibrium, when current flows, and the law of mass action is applied to a non-equilibrium situation, different from the systems considered when the law was originally derived. Non-equilibrium systems are important. Almost all of biology occurs away from equilibrium. Almost all devices of our technology function away from equilibrium.

Device design in the electrical world is helped because electrical devices obey simple laws with unchanging parameters with considerable accuracy over a wide range of conditions. Device design in the chemical world is difficult because simple laws are not obeyed in that way. Rate constants of the law of mass action are found experimentally to change from one set of conditions to another. The law of mass action is not robust in most cases and cannot serve the same role that circuit models do in our electrical technology.

One reason the law of mass action is not robust is that it is nearly always inconsistent with conservation of current. Inconsistent theories tend to need to adjust parameters (sometimes dramatically) as conditions change. Variational methods are designed to enforce consistency of physical laws. Variational methods have only recently been developed to ensure that charge flow is conserved globally, along with mass, in dissipative systems like ions in solution or proteins. The Energy Variational Approach *EnVarA* developed by Chun Liu, more than anyone else, should allow the development of more robust models of chemical, biochemical, and biological systems, making practical devices more easy to design and build.

I believe robust models and device designs in the chemical world will not be possible until continuity of current is embedded in a generalization of the law of mass action using a consistent variational model of energy and dissipation.

# **Main Paper**

Introduction. The law of mass action is used widely, nearly universally, in chemistry to describe chemical reactions. The law of mass action does not automatically conserve current, as is clear from the mathematics of a simple case, chosen to illustrate the issues involved. If current is not conserved in a theory, charges accumulate that cannot accumulate in the real world. In the real world, tiny charge accumulation—much less than one per cent—produce forces that change predictions of the theory a great deal. Indeed, in the real world tiny charge accumulation produces forces that destroy biological membranes and thus living systems, forces large enough to ionize atoms, creating the ionized plasma of electrons we call a spark, or lightning, that can destroy laboratory apparatus, if not the laboratory itself.

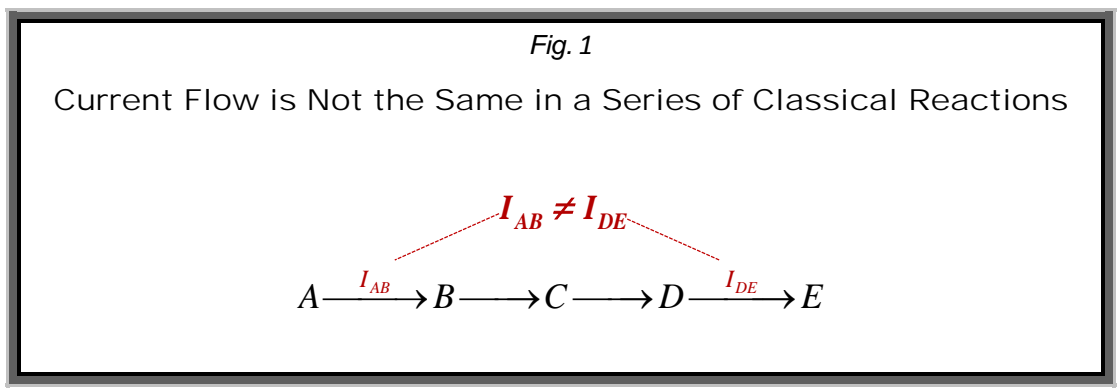

Fig. 1

**Current Flow is Not the Same in a Series of Classical Reactions**

$$I_{AB} \neq I_{DE}$$

$$A \xrightarrow{I_{AB}} B \longrightarrow C \longrightarrow D \xrightarrow{I_{DE}} E$$

Turning to mathematics for proof. Readers should distrust a claim of conflict between long established laws and so I turn to mathematics to show the conflict unambiguously, hopefully without argument. Consider the chemical reaction $X \rightleftharpoons Y \rightleftharpoons Z$: the current in the two reactions $X \rightleftharpoons Y$ and $Y \rightleftharpoons Z$ is not the same. $I_{XY} \neq I_{YZ}$. Rather, the difference in current is easily shown to be $I_{XY} - I_{YZ} = z_X k_{xy} F[X] - z_Y k_{yx} F[Y] - z_Y k_{yz} F[Y] + z_Z k_{zy} F[Z]$. The difference is not zero in general circumstances, nor robustly. The difference can be zero only for special circumstances. (Details are in eq. $(6) - (11)$ below).

The Appendix shows that the current imbalance $I_{XY} - I_{YZ}$ between the reactions quickly produces enormous forces. Enormous forces are not observed and we conclude that the law of mass action cannot be generally true, although it can be useful (and true) under special circumstances, as also shown in the Appendix, and of course in the special circumstance of equilibrium, when current does not flow.

Law of mass action is an incomplete truth. The 'law' of mass action seems to be an example of an " … incomplete truth [that] may become ingrained and taken as the whole truth … [thereby confusing] … what is *only sometimes* true with what is *always* true…" (slight reworking of Richard Feynman, p. 15-61 of ref [59]). The 'laws' of science are usually learned early in our careers before we have refereed grants and papers, before our critical skills are honed. These scientific laws are honored because of their historical role, as much as their logical importance. The laws are often residues of revolutions that once gave us new knowledge. Old ideas can have a life of their own, a momentum that is hard to change when knowledge expands It is easy to continue to use old ideas uncritically even after they have been overtaken by new knowledge.

**Law of conservation of current is different: it is universal**. Conservation of current is as universal as any law in science. It is a law that has not aged as the law of mass action has.

Maxwell's equations and conservation of current have a present role and validity at least as important as their historical one. They are said to be exact at any single time and over any distance scale that can be observed, from sub-atomic, and much smaller, to interstellar, from times much shorter than those of atomic motion to the years it takes light to travel from stars and nearby galaxies. Their daily use in high energy accelerators involving inconceivably small distances and brief times, their use in the microwave and faster devices of our technology, their use in interstellar astronomy are convincing practical proofs of the nearly universal validity of these equations of electrodynamics.

The role of Maxwell's equations depends on the subtle idea of *displacement current in a vacuum, current not carried by the movement of mass*, and this idea is easy to lose sight of in chemical applications where other forms of displacement current involving the properties of matter (usually called 'polarization' in the chemistry literature, along with quite different phenomena, e.g., charge distribution in a carbonyl bond or peptide linkage) receive more attention. The displacement current needs to be incorporated into chemical models, in my view, if they are to satisfy both conservation of mass and conservation of current. Both material displacement current accompanying polarization phenomena, and vacuum displacement current need to be incorporated. On the time scale of atomic motions $10^{-16}$ sec, both displacement currents are substantial, and must be incorporated if conservation of current is to be enforced.

**How has the law of mass action produced such useful chemistry, over so many years, if it does not conserve current flow?** This is an important question that clearly needs an answer.

The answer is that chemistry has not been interested in nonequilibrium systems of molecules with current flow, as much as it has been interested in the molecules themselves. Chemists make molecules, not currents, with the notable exception of electrochemists, and there the difficulties in dealing with long range electric fields have long been recognized [7, 78, 79].

In classical biochemistry, the law of mass action has been used in another context, more biological, with less attention to the virtues of consistency and invariant physical parameters [47]. A senior biochemist recently said "The art of biochemistry is to study enzymes in conditions that give insight to biological function, even if the rate constants do not fit results in a range of conditions. The art is to choose experimental conditions in which biochemical reactions are well behaved and rate constants resemble those in real biological systems, so results are useful in understanding how living systems work."

**The law of mass action was developed to deal with isolated systems**, originally with perfect gases [8, 82, 158, 178, 183], and allowed chemists to deal with molecular and atomic reactions before physicists were convinced that atoms existed.

But isolated chemical reactions must contact the outside world to pass signals and interact with it, just as electronic systems contact the outside world and pass signals through inputs and outputs. Biological systems contact surrounding solutions and cells through ion channels and transporters. The law of mass action was not designed to contact the outside world. It was designed to help chemists build and understand individual molecules. Signals and connections with the outside world almost always involve electricity because charge flows through the contacts that connect chemical reactions with the outside world. The contacts are usually the boundary conditions of mathematical models.

Engineering deals with systems that are not isolated. Its devices contact the world through power supplies and inputs and outputs. Devices have outputs that follow inputs according to simple rules, for example, the output of an amplifier follows the input according to the gain. The input output rules are derived from Kirchhoff's current law, i.e. continuity of current in one dimension, and Kirchhoff's voltage law.

Biology deals with systems that are not isolated. Living systems usually have inputs and outputs and are driven by concentration gradients that are power supplies. Biological systems interact with surrounding solutions, cells, and tissues.

I argue that the *law of mass action must be extended to deal with inputs and outputs and flow of electrical charge* if theories and simulations of nanodevices (technological or biological) are to be useful in more than one set of conditions.

Ionic solutions satisfy two conservation laws, conservation of mass and conservation of charge.

Chemistry uses mass conservation almost everywhere, in the form of the law of mass action [4, 122, 151] (see eq. $(6)-(11)$ below).

Physics uses conservation of charge and current whenever it deals with electricity [59, 109, 115, 152, 156, 188, 193, 210, 217]. The flow of charge is continuous in Maxwell's equations, without loss in series circuits,

$$\nabla \cdot \mathbf{I} = 0, \tag{1}$$

The current $\mathbf{I}$ is simple in a vacuum and in a vacuum that also contains particles (e.g., electrons in a vacuum tube). In the latter case the total current is what I call the Maxwell current $\mathbf{I}_{\text{Mxw}}$ because this is the invariant introduced by Maxwell to allow the light waves of his equations to propagate forever through a vacuum.

$$\mathbf{I}_{\text{Mxw}} = \mathbf{I}_{\text{particles}} + \varepsilon_0 \frac{\partial \mathbf{E}}{\partial t} \tag{2}$$

Current in matter is more complex than in a vacuum tube because it includes complex movements of charged particles (and charge inside particles) as well. The complex movement of particles must be described by coupled field equations because the movements are driven by many forces, e.g., mechanical, convection, thermal, diffusional, as well as electric. The issues involved are illustrated in Fig. 2 and the following extensive discussion. I then advocate using variational treatments of forces and movements because variational methods *automatically* enforce consistency with the field equations of all the forces.

The material displacement current accompanying polarization $\mathbf{I}_{\text{Pol}}$ is customarily treated differently from the rest of the movement of particles in a tradition started by Faraday and reinforced by Maxwell who did not recognize the existence of permanent, i.e., fixed charge.[1] The

---

[1] The existence of permanent charge as the main source of the electric field was not recognized, at least in the UK and at Cambridge, until JJ Thomson discovered the electron. JJ Thomson wrote a book [202] (long after Maxwell's death) identified " … as a sequel to Professor Clerk-Maxwell's Treatise on electricity and magnetism" (see title page of JJ Thomson's treatise). *A search of the PDF file shows the book does not contain the word charge.*

Soon after publishing his Maxwell's Treatise—without mentioning charge—JJ Thomson discovered the electron as a ray in a vacuum [204, 206]. He discovered charge in its most permanent invariant form. Each electron had a permanent invariant charge of 1*e*. After the discovery of the electron, it was clear that *permanent charge is the main source* polarization of matter is the distortion of the distribution of charge produced by the electric field (roughly analogous to the distortion of the oceans produced by the gravitational field of the moon, that produce tides on our beaches twice a day).

<u>Material displacement current</u> $\mathbf{I}_{\text{Pol}}$ can be separated from the complex movements of particles in matter by two properties.

(1) Material displacement current is transient. If an electric field is applied to matter, $\mathbf{I}_{\text{Pol}}$ flows as the spatial distribution of charge changes (i.e., as it polarizes), but eventually, flow ceases and $\mathbf{I}_{\text{Pol}} \to 0$, as $t \to \infty$ even if the local electric field is maintained constant.

(2) If an electric field is applied, and then turned off, the charge $\mathbf{Q}_{\text{Pol}} = \int_0^\infty \mathbf{I}_{\text{Pol}}\, dt$ that flows when the field is turned on is equal to the charge that flows when the field is turned off, when measured for long enough time periods.

The total displacement current $\mathbf{I}_{\text{Dis}}$ is often isolated and identified by these features in experiments involving transients. In experiments involving sinusoidal applied fields, $\mathbf{I}_{\text{Dis}}$ is usually recognized by its ninety degree phase shift.

$$\mathbf{I}_{\text{Dis}} = \varepsilon_r \varepsilon_0 \frac{\partial \mathbf{E}}{\partial t}$$

$$\underbrace{\mathbf{I}_{\text{Dis}}}_{\substack{\textbf{Total} \\ \text{Displacement} \\ \text{Current}}} = \underbrace{\varepsilon_0 \frac{\partial \mathbf{E}}{\partial t}}_{\substack{\textbf{Vacuum} \\ \text{Displacement} \\ \text{Current}}} + \underbrace{\mathbf{I}_{\text{Pol}}}_{\substack{\textbf{Material} \\ \text{Displacement} \\ \text{Current}}} ; \qquad \underbrace{\mathbf{I}_{\text{Pol}}}_{\substack{\textbf{Material} \\ \text{Displacement} \\ \text{Current}}} = \underbrace{\left(\varepsilon_r - 1\right)}_{\text{Susceptibility}} \varepsilon_0 \frac{\partial \mathbf{E}}{\partial t} \qquad (3)$$

The vacuum displacement current is a large fraction (say half) of the displacement current on the time scale of atomic motion, faster than $10^{-15}$ sec.

---

of the electric field. Induced charge (the field dependent component of what chemists call 'polarization') was seen to be a consequence of the field, to be computed as an output of a model of the field, produced by the displacement of matter, more or less akin to gravity's displacement of the oceans that produces tides. The induced charge contributes to the resulting field, to be sure, but it is not the main source, any more than the tides are the main source of the gravitational interactions of earth and moon.

Maxwell, Faraday, and JJ Thomson's 'sequel to Maxwell' had the opposite view. The pre-electron JJ Thomson, along with Faraday, and probably Maxwell, viewed induced charge somewhat mysteriously (in my opinion), as almost the same thing as the electric field and in some sense as its source. Physicists today give permanent charge a prominent place and view the *relation of charge and field is the essence of electrodynamics,* as one imagines JJ Thomson did after his discovery of the electron. The many textbooks that follow Abraham and Becker [1] document this view.

All current produces a magnetic field, according to Maxwell's extension of Ampere's law

$$\nabla \times \mathbf{B} \quad = \mu_0 \mathbf{I} \quad = \mu_0 \left( \mathbf{I}_{\text{Mxw}} + \mathbf{I}_{\text{Pol}} \right) = \mu_0 \left( \mathbf{I}_{\text{particles}} + \mathbf{I}_{\text{Dis}} \right)$$

$$= \mu_0 \left( \mathbf{I}_{\text{particles}} + \overbrace{\varepsilon_0 \frac{\partial \mathbf{E}}{\partial t}}^{\substack{\text{Vacuum Displacement} \\ \text{Current}}} + \overbrace{\left( \varepsilon_r - 1 \right) \varepsilon_0 \frac{\partial \mathbf{E}}{\partial t}}^{\text{Susceptibility}} \right) \quad (4)$$

The divergence operator $\nabla \cdot \mathbf{f}$ evaluates the conservation of flow in the vector field $\mathbf{f}$ it acts on (as its derivation from integral relations shows nicely [190])

$$\nabla \cdot \left( \nabla \times \mathbf{B} \right) \quad = \nabla \cdot \left( \mu_0 \mathbf{I}_{\text{particles}} + \mu_0 \varepsilon_0 \frac{\partial \mathbf{E}}{\partial t} + \left( \varepsilon_r - 1 \right) \mu_0 \varepsilon_0 \frac{\partial \mathbf{E}}{\partial t} \right) |$$

$$= \mu_0 \left( \nabla \cdot \mathbf{I}_{\text{Mxw}} + \nabla \cdot \mathbf{I}_{\text{Pol}} \right) = 0 \quad (5)$$

so $\qquad\qquad\qquad \nabla \cdot \mathbf{I} = 0$

where we use the vector identity that the divergence of a curl is always zero. $\mathbf{B}$ is the magnetic vector field; $\mu_0$ is the magnetic constant, the magnetic 'permeability' of a vacuum; $\varepsilon_0$ is the corresponding 'electrostatic constant', the permittivity of free space. Note that $\mu_0 \varepsilon_0 = c^{-2}$ where $c$ is the velocity of light. Magnetism only arises from current flow. It is mysterious that magnetic charges (monopoles) do not exist, i.e., $\nabla \cdot \mathbf{B} = 0$.

Maxwell's equations (1) & (5) guarantee that current is exactly the same everywhere in a series of two terminal devices. If parameters or geometry are changed, electrical forces and potentials change automatically to ensure the same current flows everywhere (in a series system involving Maxwell's equations). Interruption of current anywhere in a series circuit interrupts current flow everywhere, even signals carried by currents far away from the interruption.

The law of mass action needs to deal with displacement current so it can enforce global continuity of current flow. If a particular chemical reaction is at equilibrium, and no current flows, conservation of charge can be enforced easily on the law of mass action. But current flow presents a different situation because the law of mass action must then be modified to deal with the global properties of the electric field. If we are dealing with nanodevices (technological or biology) chemical reactions are connected to the outside world and the law of mass action must be extended to enforce continuity of current flow everywhere under all conditions.

Biological systems involve both chemical reactions and charge. Biological systems are always embedded in ionic solutions, and nearly always involve chemical reactants and enzymes with electrical charge, even if (like water) their net charge is zero. Substrates of reactions catalyzed by enzymes are usually charged and are nearly always dissolved in complex solutions containing the ions $Na^+$, $K^+$, and $Cl^-$ and often $Ca^{2+}$. Trace concentrations of calcium are often used by biological systems as controls that turn biological function on or off. Reactants are almost always charged in biological systems. Reactants almost always flow and carry current in biological systems. Equilibrium and death are nearly synonymous in life.

Biological systems must satisfy conservation of current along with conservation of matter. We face a problem when we try to apply both conservation laws together: The Law of Mass Action does not conserve current.

'The Law of Mass Action does not conserve current' seems an unlikely statement and so mathematical proof is needed more than verbal argument. We examine a simple sequence of chemical reactions and ask "Does the law of mass action conserve current?" Is current the same everywhere in the series of reactions? Do the potentials change automatically so current is always the same everywhere in a series circuit?

The same questions can then be asked of whatever series of reactions are of interest. Sometimes the answer will be that current does not flow or does not matter and the law of mass action can be easily modified to conserve charge. In some symmetrical reactions, the answer will be that current and mass are both conserved, as shown in the Appendix. More often, the answer will be that current and mass are not both conserved, as shown in more detail in the Appendix. In that case, the law of mass action must be extended to maintain continuity of current.

Eq. (6) shows the reactions we use to illustrate the problem

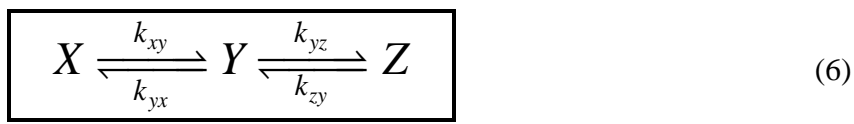

$$(6)$$

The reactions of eq. (6) were chosen to show the problem in the simplest case. We define 'law of mass action' for this paper as just eq. (6) & (9) with symbols defined below. The rate constants in eq. (6) & (9) are taken as constants and thus of course are uncorrelated and independent of each other. If the law were robust, the rate constants found experimentally under one set of conditions would be found under another set of conditions. Such is sometimes the case, but not very often [47, 48].

Generalizations of rate constants are sometimes made but, as discussed below—in the section 'How to extend the law of mass action?'—generalizations in the literature [4, 122] deal with 'chemical correlations'[4] and non-ideality of some types, but they do not deal with current flow, to the best of my knowledge. They do not allow the law of mass action defined here by eq. (6) & (9) to globally satisfy conservation of current eq. (1). The generalizations are well designed to deal with systems close to equilibrium with some types of non-ideality, but the generalizations do not discuss current flow, and have little[3] or no[4] discussion of the nonidealities produced by the ionic atmosphere (in the equilibrium case [23]), and changes of shape of the ionic atmosphere (in the nonequilibrium case [37, 67-69, 116, 127, 181, 207]).

Current flow is important in most applications of the law of mass action. It is almost always present if the reaction is part of a device that communicates with the outside world. The global nature of the electric field (illustrated in Fig. 1) allows remote devices and boundary conditions to change local atomic flows, an effect not necessarily present at equilibrium. In a nonequilibrium system, electrical forces and potentials change everywhere—automatically as a result of the equations of the electric field—to ensure the same current flows in all places (in a series system). Indeed, interruption of current in a series of reactions stops current anywhere, even far away.

Of course, there are special cases in which mass action can by itself conserve current flow exactly and those in fact may be the cases where it has proven most quantitatively useful, particularly when extended to deal with nonideality [4, 122]. In addition, reactions may conserve current approximately. In other cases, like eq.(6), the reaction will not conserve current, even approximately. Each case needs to be studied separately. The Appendix provides more detail.

The central fact—that applies to any chemical reaction, not just eq. (6) & (9)—is that the global realities of the electric field need to be embedded in the atomic scale treatment of the reaction, particularly when current flows.

## Proof

The current flow in the reactions of eq. (6) is easily shown to be

$$I_{XY} = Fz_X \cdot k_{xy} [\![ X ]\!] - Fz_Y \cdot k_{yx} [\![ Y ]\!]$$

$$I_{YZ} = Fz_Y \cdot k_{yz} [\![ Y ]\!] - Fz_Z \cdot k_{zy} [\![ Z ]\!]$$

(7)

Units for net current[†] $I_{XY}$ are (cou/liter)/sec = cou/(liter sec). Unidirectional flux $J_{xy}$ are (moles/liter)/(sec) = moles/(liter sec). Double brackets like $[\![ X ]\!]$ indicate activities, the generalization of concentration (number density) needed in biological solutions, as discussed below. Units for rate constants are $\big( \text{moles/(liter sec)} \big) \big/ \big( \text{moles/liter} \big)$ = 1/sec . The valences (i.e., charges on one molecule) of each reactant are $z_X$ or $z_Y$ . $F$ is Faraday's constant.

In general, the conservation of current law eq. (5) is violated:

$$\boxed{I_{XY} \neq I_{YZ} \text{ , for a range of concentrations, rate constants, or charges.}}$$

(8)

The main result of eq. (7) & (8) and the Appendix is simple: the law of mass action itself conserves current only under special symmetric circumstances. The law of mass action does not automatically change the electrical potentials to ensure that the current flow in a series circuit are the same everywhere.

<u>Details of Proof</u>: Rate equations equivalent to chemical reactions involve *net flux* like $\overline{J}_{XY}$ . The chemical reactions in eq. (6) defines the net fluxes, units moles/(liter sec),

$$\overline{J}_{XY} = k_{xy} [\![ X ]\!] - k_{yx} [\![ Y ]\!]$$

$$\overline{J}_{YZ} = k_{yz} [\![ Y ]\!] - k_{zy} [\![ Z ]\!]$$

(9)

with the definitions of rate constant $k_{xy}$ , for example

---

[†] Unidirectional fluxes like $J_{xy}$ use lower case subscripts. Net fluxes like $\overline{J}_{XY}$ use upper case subscripts and are marked with an over-bar. Currents like $I_{XY}$ use upper case subscripts because they are net currents, and are not unidirectional. Unidirectional currents are not defined or used in this paper.

$$k_{xy} = \frac{J_{xy}}{[\![X]\!]} = \frac{J_{xy}}{-\dfrac{d}{dt}[\![X]\!]_{J_{yx}=0}} \qquad (10)$$

**Remark.** Unidirectional fluxes like $J_{xy}$ are conventionally measured by tracers—originally radioactive isotopes, now usually fluorescent probes—flowing *into* an acceptor solution in which the concentration of tracer is zero. The concentration in the acceptor solution of the substance is not zero in most cases, although the concentration of the tracer is (nearly) zero. Note that the activity $[\![X]\!]$ depends on the concentration of *all of the other ions* in a significant way in nonideal solutions [10, 16, 30, 41, 43, 45, 47, 58, 63-65, 70, 71, 110, 121, 123-125, 127, 129, 130, 141, 144, 154, 174, 180, 211, 212, 218]. In nonideal solutions, changes in the concentration of one substance $[Y]$ change the activity of another substance $[\![X]\!]$ and also change its flow.

**Back to the proof.** The proof of eq. (7) is completed by writing net fluxes like $\bar{J}_{XY} = J_{xy} - J_{yx}$ as the difference of unidirectional fluxes. Net flux is converted into net current like $I_{XY}$ using the proportionality constant $zF$ (with $z$ for the species in question) between flux and current.

$$I_{XY} = Fz_X \cdot k_{xy} [\![X]\!] - Fz_Y \cdot k_{yx} [\![Y]\!]$$

$$I_{YZ} = Fz_Y \cdot k_{yz} [\![Y]\!] - Fz_Z \cdot k_{zy} [\![Z]\!]$$

`(11)

**Obviously, $I_{XY} \neq I_{YZ}$**

*This ends the proof.*

**Organization of paper.** It seems unlikely that conservation of current and conservation of mass are in conflict in the law of mass action, so mathematics is needed to show that current is not conserved. I treat a simple but widely used case, hoping that this will be more convincing than words and easier to extend to other reaction schemes. *The methods used can be applied to any series of reactions to see if they satisfy both conservation of current and conservation of mass.* In general, an additional constraint and equation is needed to enforce the global continuity of current flow. The additional constraint is introduced in the special case that the chemical reaction describes spatial movement through narrow channels. In that case, a fully consistent treatment is possible using a variety of models of charge movement. The general case where chemical reactions describe covalent bond changes is not worked out in detail here. In my view, the general case must be dealt with by variational methods that enforce consistency of all the physical laws. Energetic variational methods appropriate for this purpose (in dissipative systems like ionic solutions) are described.

The Appendix shows that charge imbalance predicted by the law of mass action would likely have noticeable effects. A small charge imbalance quickly produces potentials and forces that destroy membranes (~0.3 volts), molecules in a liquid (~2 volts, $H_2O$, with Ag|AgCl electrode), molecules in a gas (~480 volts for $H_2O$ gas), atoms in a gas (~14 eV for $N_2$). These effects would destroy experiments and are not found in the laboratory, which would also soon be destroyed. We conclude that theories that predict charge imbalance because they do not conserve current flow are not acceptable approximations to the world of experiments or life.

# Current Flow in Complex systems

We now discuss current flow in some detail. As we discuss current flow through salt water, vacuum capacitors, dielectric capacitors, vacuum tube diodes, semiconductor diodes, resistors, wires and batteries we will see the bewildering number of ways that the electric field changes to ensure that current flow is always continuous.

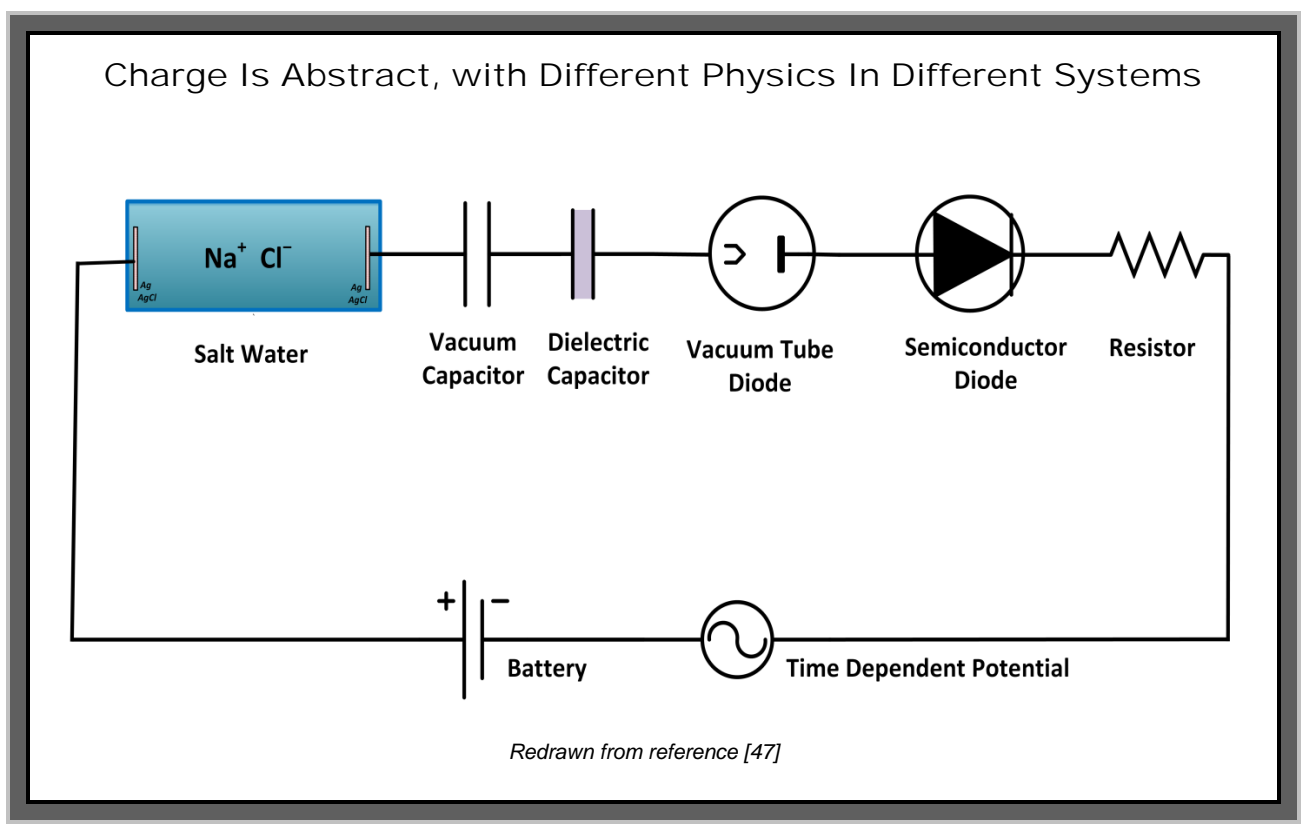

**Fig. 2.** We see that charge in one physical system is quite different from charge in another. Charge flow is **not** simply the physical movement of particles of definite charge (and mass). Charge flow in a vacuum capacitor, in an ionic solution, in a wire are all quite different. Current is not just the movement of ions or electrons or protons. Note that the electric charge in different devices varies on very different scales from subatomic (in dielectrics) to atomic (in diodes, ionic conductors, etc.) to macroscopic. Thus, *treatments of charge must be multiscale.*

<u>Charge is an abstraction that is conserved</u>. Physicists teach that conservation of charge and current are universally true, exact from very small to very large scales and very small times to very large times.[2] What physicists often do not teach clearly is that charge is an abstract idea.

---

[2] Even the Casimir effect of quantum physics [177] is seen as a property of Maxwell's equations by some [111].

Many students believe that charge is always carried by particles and so conservation of mass (of charged particles) implies conservation of charge and current. That is not true. Charge is carried between the plates of a vacuum capacitor as a displacement current that does not involve the movement of particles [59, 115, 188, 217]. Current flow in semiconductors is carried by mathematical fictions called quasi-particles that move according to classical physics [109, 152, 156, 193, 210], not involving solutions of Schrödinger's equation. This is an important property of semiconductors [117, 120], crucial to their use.

An essential idea of electricity and magnetism as explained in textbooks [115, 188, 217] is that *charge flow is continuous* (without loss in a series circuit) __no matter what the physical nature of the charge__. Electrical potentials and forces change automatically to guarantee continuity of current flow under all conditions, in experiments and in the equations of electromagnetism. I found Saslow's [188] treatment of continuity of current particularly clear and useful.

Fig. 2 tries to show this idea in a concrete way that is easy to build in a lab. 'Completing the circuit' implies that current in every device is the same. Continuity of current—the same as Kirchhoff's current law in a one dimensional sequence of reactions—says that time-varying currents are the same in any series of devices even if they have very different physics and different constitutive laws, even if they involve chemical reactions (Fig. 1 and eq.(6)).

Ionic Conductor. The ionic conductor in Fig. 2 is a cylinder containing $Na^+Cl^-$. Here, current flow (at a frequency say of 1 Hz in a $2 \times 10^{-2} M$ solution) is almost entirely the physical movement of charged particles, of ions, say sodium and chloride ions and follows simple constitutive laws (when concentrations are $< 2 \times 10^{-2} M$ and flows are not large enough and do not last long enough to change concentrations). These ions are hard spheres. The finite size of these spheres is significant in the ionic mixtures found everywhere in biology and in general at concentrations greater than say $2 \times 10^{-2} M$. The finite size makes constitutive equations (valid at all concentrations of mixtures of different types of ions) much more difficult than the classical constitutive equations for quasi-particles that are points. Finite size implies saturation effects—space cannot be filled more than once—and these imply that 'everything interacts with everything else'.

Numerical difficulties in dealing with spheres are substantial. *Spheres must be computed in three dimensions because spheres do not exist in one and two dimensions.* That is to say, objects with a single radius have very different surface to volume ratios in one, two and three dimensions and so fill space very differently. Phenomena in which spheres fill a significant fraction of three dimensional space are not easily approximated in one or two dimensions. Computation of the forces that prevent overlap of spheres is difficult because those forces are strong and vary steeply in three dimensions. Bottom line: nonideal solutions remain a challenge, as documented below.

Vacuum Capacitor. We move to a vacuum capacitor, in which the space between the two plates is completely empty of matter (as it would be in outer space, for example). The current flow through this capacitor is just as real as the movement of ions of $Na^+Cl^-$ in the cylinder even though no particles or spheres are present, and no mass moves at all.

The displacement current between the plates of the vacuum capacitor is a property of the electric field itself, as explained in textbooks of electricity and magnetism [59, 115, 188, 217] and is described by the exact and simple constitutive equation for the vacuum $i_{displacement} = C(\partial V / \partial t)$, where the displacement current $i_{displacement}$ (amps) is strictly proportional to the capacitance $C$ (farads) and the time rate of change $\partial V / \partial t$ of the voltage across the vacuum capacitor. Unlike other constitutive equations, the constitutive equation for vacuum current is exact, valid to some eighteen significant figures [217].

This displacement current induces a magnetic field just as current carried by ions produces a magnetic field. Indeed, without displacement current in a vacuum, Maxwell's equations do not allow sunlight to propagate through the vacuum of space. With this exact expression for displacement current, light propagation is a solution of the Maxwell equations, and in fact the speed $c$ of propagation of light can be computed from measurements of electrical and magnetic constants, entirely independent of measurements on light itself, by Maxwell's remarkable formula $c = 1/\sqrt{\mu_0 \varepsilon_0}$. Light propagates according to Maxwell's equations over astronomical distances, so we know that the constitutive equation that calculates that speed must be accurate to many significant figures.

<u>Dielectric Capacitor</u>. In the dielectric capacitor in Fig. 2 (filled with real material, for example the plastic = Polytetrafluoroethylene = PTFE = Teflon), current flow is more complex, and involves the effect of an applied electric field on the spatial distribution of the electric charge intrinsic to the atoms, molecules, and substance of the dielectric. ('Intrinsic' here means the distribution of charge present when there is no applied electric field.) Note that the electric charge in a dielectric, or an ionic solution, or a protein or nucleic acid, for that matter, varies significantly on all scales from subatomic to macroscopic.

The properties of dielectrics cannot be described in detail here because they vary so much with material, voltage, and time. But some properties need emphasis. Intrinsic charges do **not** move from plate to plate in a real capacitor. The current that flows from plate to plate and within the dielectric is a dielectric displacement current not carried by the movement of mass (any substantial distance). The current in a real capacitor is an abstraction. It is the sum of the vacuum displacement current and the material displacement current $\mathbf{I}_{\text{Pol}}$ (in the dielectric) produced by the distortion of intrinsic charges. The material displacement current describes the movement of a *charge on a nonlinear time dependent spring (with damping) that eventually stops moving* after being perturbed by an external electric field for a finite time. Calculating this material displacement current involves solving a nonequilibrium time dependent version of the Schrödinger equation, a difficult task, involving a macroscopic number of atoms and millisecond time scales in cases of biological interest.

<u>Ideal dielectrics do not approximate ionic solutions</u>. Scientists have avoided the difficulties of solving the Schrödinger equation in a macroscopic system like a dielectric, or ionic solution, by using approximations. They discuss an ideal dielectric with properties independent of field strength and independent of time. The approximation over a wide range of electric field strength is satisfactory in most materials.

The approximation of a time independent ideal dielectric is poor over a wide range of times. For example, most solutions of ions in water need effective dielectric coefficients to describe the polarization charge induced by permanent charges. This dielectric charge changes—after a step electric field is applied—from about 2 to about 80 as time moves on, from zero, to $10^{-15}$ sec to say $10^{-5}$ sec and the change in dielectric charge depends on the substance.[9] There is not a universal constitutive law or approximation for the time dependence of dielectric properties.

A factor of 40 change with time is not small. Multiscale analysis must deal with this multiscale problem because atomic motion has macroscopic consequences on all time scales.

The time dependence of real dielectrics needs to be dealt with in biological applications: ideal dielectrics do not approximate the properties of ionic solutions in which biology occurs. The time dependence of polarization charge in proteins has been extensively studied [167-169] and varies over the whole time scale from atomic to macroscopic. It often has slow components, $10^{-3}$ sec or slower.

Biology is controlled by atomic structures that move significantly in $10^{-15}$ sec but have dramatic effects on biological functions $10^{-5}$ sec to 10 sec later. The time dependence of these induced polarization charges appear in simulations (with atomic resolution) as changes in the orientation and induced polarization of water molecules, and the distribution of ions, as well as the distribution in proteins of permanent charges, permanent dipoles, and induced dipoles.

Atomic scale simulations must last a long time. Atomic scale simulations seeking agreement with experimental data—available for a wide range of solutions [9], and proteins [167-169]—must last long enough to account for the discrete charge movements that produce an effective dielectric coefficient of 80, as measured experimentally. Otherwise, substantial changes in the electric field will not be seen in the simulation, even if those changes in the electric field have great biological significance.

Electric fields in biology are strong. The nerve signal is a change in the electric field of the greatest biological significance. Information transfer in the nervous system, coordination of contraction in muscle, including the coordination of cardiac contraction that allows the heart to pump blood, are all directly controlled by the action potential, the biological name for the nerve signal.

Electric fields during a biological action potential are slow (say $10^{-3}$ sec), and macroscopic in scale, spreading $10^{-3}$ meters, propagating one meter (in humans), and they are remarkably multiscale. Handfuls of atoms in a channel protein control the macroscopic propagation and the nerve signal itself is current carried by ions (that are single atoms) moving through those proteins.

The electric field of the action potential is strong, typically some 0.1 volts across a $2\times10^{-9}$ meter thick membrane or a $3\times10^{-10}$ meter selectivity filter (EEEE group in a calcium channel), $5\times10^{7}$ to $3\times10^{8}$ volts/meter. These electric fields distort the intrinsic distribution of charge within the dielectric on many scales of time and distance, including electrons inside molecules and atoms, They reorient polar molecules that have an intrinsic asymmetric distribution of charge. Charges only move a small amount in dielectrics in response to the applied electric field——reminiscent of the sloshing of tides in the ocean and on beaches on the earth created by the moon's gravitational field—*and they eventually return to their resting position when the field is turned off* but those small movements of charge produce large effects because the electric field is so strong.

Dielectric currents $\mathbf{I}_{Pol}$ share some of the properties of vacuum displacement currents but they do not have a universal exact constitutive law, not even approximately. It is important not to confuse dielectric $\mathbf{I}_{Pol}$ and vacuum displacement currents $\varepsilon_0\,\partial\mathbf{E}/\partial t$. It is important to realize dielectric displacement currents can be large and produce large effects in biology, as they do in semiconductors.

Nonlinear components of dielectric current are important in biology. Nonlinear components of dielectric displacement current have large effects important in biology (e.g., in muscle [191], where they were first discovered, and nerve [3]) where they control the opening and closing of the channels that produce the action potential [209] and some enzyme function [126], as well. These nonlinear dielectric (i.e., displacement) currents can be recorded as 'gating' currents because of the Shockley-Ramo theorem [166], which is a restatement of continuity of current. The dielectric currents involve a tiny fraction of all the charges in and near an ionic channel, far less than one per cent (see Appendix, "Size of Effects", near eq.(28)), but continuity of current and the strength of the electric field make gating currents measurable, easily in the hands of skilled experimentalists.

<u>Any electric field is extraordinarily strong compared to diffusion</u>. Tiny displacement currents can be recorded routinely in many laboratories because the electric field is so strong. If it were not so strong, these tiny concerted movements of atoms would be lost amongst the Brownian thermal motion. Extraordinarily small changes in net charge are enough to guarantee that currents are the same in all elements of a series circuit because the capacitances (ratio of charge to potential) involved are often some $10^{-17}$ farads, see Appendix. Continuity of current flow is guaranteed by changes in electrical forces and potential resulting from very very small amounts of charge.

<u>Concentration effects are very small</u>. On the other hand, a one percent deviation in density of mass has a tiny effect on diffusion and chemical reactions, hardly noticeable. Diffusion forces are tiny perturbations on the electric force field and energies involved in diffusion are tiny perturbations in the total energy of charged systems. Electric forces and energies are not small parts of the total energy and it is not wise—to say the least—to treat them as perturbations of uncharged systems. Small changes in charge distribution produce large changes in flow because flow tends to be an exponential function of the spatial profile (even when the profile does not have a large, single, symmetrical peak [40] of potential energy) and the profile of energy is a sensitive function of permanent and dielectric charge, ionic conditions, etc [52]. It is not wise [54] to neglect the effect of permanent charges of the channel protein on the shape and size of the electric field as is done in many continuum [20, 25-28, 31-33, 75, 77, 90, 131-140, 182, 198] and rate models of gating and permeation of ion channels, and most treatments of enzyme function and catalysis.

<u>Charge is indeed an abstraction</u>. In each of the devices in Fig. 2, charge follows different laws, because it has different physical properties, sometimes carried by charged particles, sometimes produced by the rate of change of electric field (displacement current), sometimes the 'movement' of quasi-particles, sometimes the 'movement' of electrons in a macroscopically delocalized quantum state of a wire.

<u>Charge always flows without loss in each device</u>. That abstract property of charge movement is always true, because the potentials change automatically to ensure continuity of current. But the 'laws' describing current as a function(al) of time and potential (for example) depend on the physical nature of the charge and its movement. Different devices have different relationships between current, voltage, and time. These different 'constitutive' equations are described in the engineering literature in great detail.

The different constitutive equations combine with Kirchhoff's current law to describe current flow from one place and one device to another. Together, the equations describe the universal fact that interrupting current flow in one chemical reaction (of a series of reactions) will interrupt current flow in every other reaction (in that series) even if the interruption is meters away from a chemical reaction being studied on the atomic scale (scale = 1 Å). Together the equations describe the experimental fact that electric fields, forces, and potentials automatically rearrange so that Kirchhoff's current law and continuity of current flow are always present exactly on all scales no matter what is 'carrying' the current. Indeed, *the physics of current flow changes to accommodate the continuity of current*.

One is reminded of the power of the electric field when one unplugs a computer. Interruption of a circuit meters away from the diodes of the computer's power supply stops the flow of quasi-particles—holes and semi-electrons—across atomic scale junctions of semiconductor diodes, often of the **PN** variety. The electric potential changes so strongly in response to the interruption (because the power input of the computer power supply stores a great deal of charge) that the electric field exceeds the dielectric strength of air. The *automatic change in the electric field—needed to maintain continuity of current flow—is enough to 'change the physics' of the system.* Electrons are stripped off the atoms of the air, a plasma is created, a spark. We should be frightened of sparks and their electric field. Sparks start fires, sometimes in upholstery or drapes.

<u>Vacuum tubes</u>. In the next device in Fig. 2, we consider charge movement in a vacuum tube diode. Vacuum tubes control the flow of electron current by changes in their internal electric fields and were called 'valves' in the UK for that reason. Vacuum tubes, semiconductors and even some open ion channels follow simple constitutive laws of rectifiers, as described in textbooks of electronic devices or ion channels.

In vacuum tubes, current is indeed carried by a stream of isolated charged particles, electrons with a definite mass and charge, moving through a vacuum, interacting only through their electric fields. At 1 Hz, essentially all the current in a vacuum diode is carried this way. Current through a diode is not proportional to the voltage across the diode because the electric fields within the diode change shape, despite the simple physics of conduction. The fields change shape as the voltage across the diode is changed because of screening and shielding. The electric fields within the tube are different at small and large potentials. The different internal electrical fields change the current flow, creating rectification. The electric field creates a large barrier in one direction so current in that direction is small; the electric field creates a small barrier in the other direction, and current in that other direction is large.

Rectification is of historical interest, because vacuum tubes allowed the early detection of radio waves in American homes in the 1920's, as valves did in the UK. The demand for portable radios led to solid-state diodes, then solid-state 'triodes', field effect transistors, integrated circuits, and our modern digital world [179].

Vacuum tube diodes had certain difficulties—they were big, $10^{-2}$ m at their smallest, hot, greedy consumers of power, cost about \$3 in the 1950's, and electrically unstable: they drift continuously. They were quickly replaced with semiconductor diodes that cost $< 10^{-10}$ dollars each, do not drift significantly, and can be as small as $10^{-8}$ meters nowadays.

We turn now to the semiconductor diode that operates on a very different scale from the vacuum diode. The rectification of both diodes—that is their function—depends on essentially the same physics, namely the shape of the electric field *and its change in shape with the direction of current flow.* The electric field of charge must be respected on both the scale of vacuum diodes and that of semiconductor—$10^{-2}$ meters of vacuum tubes, $10^{-9}$ meters of semiconductor diodes—indeed on all scales because the electric field has such large effects on all scales. The immediate implication is that theory and simulation must calculate the electric field on all scales, using an explicitly multiscale analysis, since it seems unlikely that any one type of simulation or theory can span atomic to biological to laboratory distances and times, let alone the interstellar scales on which the laws of electricity are known to be valid.

<u>Semiconductor diodes</u>. Current in semiconductors is carried by quasi-particles [120], called holes and 'electrons' (better named quasi-electrons, or semi-electrons in my view) These quasi-electrons and holes are defined because they interact much more simply than the totality of real electrons and lattice of atoms in the semiconductor [120, 173, 175, 196, 210]. They simplify the quantum mechanical many-body problem into the much simpler motion of imagined quasi-particles.

Current is carried in semiconductors by rearrangements of charge in the entire lattice of atoms that make up the semiconductor. Fortunately for our technology and daily life, current in germanium and silicon can be approximated by classical theories that deal with one quasi-electron or one quasi-hole at a time as they move in mean fields, without requiring solution of the Schrödinger equation at all [109, 152, 156, 193, 210].

As a textbook puts it eloquently (p. 68 of [117]): "Electron is a quasi-particle consisting of a real electron and an exchange correlation hole … a cloud of effective charge of opposite sign due to exchange and correlation effects arising from interactions with all other electrons of the system." "Hole is a quasi-particle, like the electron, but of opposite charge; it corresponds to the absence of an electron from a single particle state that lies just below the Fermi level." The motion of these quasi-particles are described by mean field models, evaluated both by simulations [109, 210] and theory [109, 152, 156, 193, 210]. For example, the Poisson drift diffusion equations [113], often called *PNP* (for Poisson Nernst Planck) in biophysics and nanotechnology [12, 16, 21, 24, 41, 51, 53, 99, 214].

*PNP* is of particular importance because it is used widely, nearly universally, to design and understand the devices of our semiconductor technology, from transistors to computer chips. *PNP* is used rather widely in nanotechnology these days and increasingly in membrane biophysics and mathematical biology and electrochemistry, including battery technology, as well as the technology of cement.

Rectification in biological membranes. Early treatments of rectification in biological membranes, then called ionic conductances, ([75]; Appendix of [95]) in fact drew heavily on Mott's (nearly) contemporaneous treatments of rectification in diodes [162], but none of these were consistent: they assumed the electric field, instead of computing it. The charge in the system did not produce the assumed electric field if substituted into Poisson's equation. Goldman [75], then a graduate student of K.S. Cole, recognized the problem, but did not know how to remedy it, nor did these workers [75, 95] understand that electric fields in diodes or open channels were not constant in any sense, including space (numerous personal communications to me from K.S. Cole, 1960-1962, and A.L. Hodgkin, 1960-1995). Mott soon realized [163] that the change in shape of electric fields was a crucial phenomenon in semiconductor diodes, updating the constant field assumption of his original paper [162]. Biophysicists were evidently not aware of the evolution of Mott's understanding [89].

Semiconductor physicists understood that consistent treatments produced electric fields that varied greatly with conditions, as the fields change to force continuity of current [195]. *Indeed, transistor design would not be possible* [179, 197, 208, 210] *if computational electronics had assumed constant fields the way some physicists did early on* [162], but not for long [163].

In biophysics the importance of computing the fields from the charges (so the fields and charges were consistent) was not realized (as far as I know) until much later [51-53]. Until then, and even after, Nernst Planck equations were used in biophysics *without including the ionized acid and base groups of the channel protein* (i.e., without including the permanent charge of the protein) [20, 25-28, 31-33, 77, 89, 131-140, 182, 198]. These ionized groups make polymers into ion exchangers [83]. The equivalent doping profiles make semiconductors into devices. Leaving out ionized groups "is like studying a galaxy without stars", as a prominent physical chemist once told me. The distribution of permanent charge creates a baseline electric field. Change of shape of the electric field with concentrations, amino acid composition of the channel protein, and so on are responsible for many of the important properties of these systems. Unfortunately, the importance of changes of shape of the electric field, and of the fixed charges involved, were and still are often ignored.

<u>Charge carriers in semiconductors can be really quite strange</u> from a physical point of view. Charge carriers in silicon and germanium do not exist outside the lattice of the semiconductor as distinct entities. They do not exist in the same sense that $Na^+$ and $Cl^-$ ions exist, but are mathematical representations, with lifetimes sometimes as short as milliseconds. Quasi-particles are the second derivative of a Fermi surface of silicon and germanium semiconductors, under particular conditions. They are defined, as mentioned above, to allow easy classical analysis compared to the intractable quantum mechanical many-body problem of a macroscopic semiconductor.

Much of the success of our semiconductor, digital, and video technology is due to the accuracy of the constitutive *PNP* equations describing holes and semi-electrons. They robustly describe the characteristics of semiconductor devices of many different types, *with very different input output relations*, as different as an exponentiator and a logarithmic convertor.

*PNP* is so useful because almost all the devices of our digital technology work in a restricted set of conditions in which flows are crucial but are of the special type well described by quasi-particles moving in a mean field. Treatment of distortions of the electric field of all the atoms in the macroscopic device is not needed.

Almost all of the devices of electronics use power supplies to maintain different voltages at different locations far from the *PN* junctions of the device itself (and thus require a global treatment of the electric field). These voltages perturb the distribution of velocities of charged particles so the distribution has net flow [40, 56]. The slight perturbation is enough to imply the *PNP* equations and those provide enough nonlinearity to make amplifiers, switches, and the full set of logical circuits necessary to make a computer [173, 175, 196, 210]. In fact, in this case the *PNP* equations can be solved analytically and exactly to give intuitive pleasingly simple formulas for current flow, once the shape of the electric field is known [40].

<u>PNP equations describe a wide variety of current voltage relations and devices</u>. The *PNP* equations are the constitutive laws used to describe semiconductor devices as their current voltage relations change drastically (with voltage, for example) from that of a linear amplifier, to a switch, to an exponentiator, a multiplier or even a logarithmic converter [173, 175, 196, 210]. Nonlinear input output relations as varied as these enable a rich variety of devices.

Nonlinear input output relations as diverse as these are not described easily—or described at all, for that matter—in most areas of physics and chemistry. All the nonlinear devices in a computer are actually mathematical solutions of the *PNP* equations in a complex silicon structure built to have the particular spatial distribution of permanent charge ('doping') that produces the desired properties of the device, i.e., the input output relations.

The predictive power of *PNP* is very important in the design of robust semiconductor devices—that do not fail even when used many millions of times a second in computers that contain a trillion transistors. In fact, *the intrinsic physical properties of semiconductors are adjusted by their designers so PNP remains a good description [29, 34, 35, 185], even as the size of the device has been dramatically decreased.* The concentration of fixed charge dopants, geometry, and recently even dielectric coefficients are adjusted by semiconductor engineers in their successive iterations of Moore's law [153, 159, 160], so *PNP* remains a good description as performance is increased by factors of a billion or so, over 50 years.

Evidently, *reliable design is more important than raw performance*. It seems more important for the designers (and marketplace) that an equation describes behavior robustly and accurately over a range of conditions than that the device be as fast or small as possible [29, 34, 35, 185]. Evolutionary selection in biology also seems to choose robustness over efficiency in many cases. Devices of nanotechnology need to be similarly robust, I believe, before they will be used extensively.

**PNP is not enough, however, when ionic solutions are involved**. A great deal of effort has been spent applying **PNP** equations to electrochemical systems [12, 16, 21, 24, 41, 51, 53, 99, 214] hoping they might serve as adequate robust constitutive equations, but that is not the case. The nonideality of ionic solutions, arising in large measure from saturation effects produced by the finite size of ions, demands more powerful mathematics than the partial differential equations of **PNP** used in computational electronics.

Resistor. The next device we discuss in Fig. 2 is a resistor, which in some ways is the easiest to describe, because current is proportional to voltage with a single proportionality constant over a wide range of voltages, times, and conditions. A resistor follows Ohm's constitutive law with a resistance independent of potential over a wide range from $10^{-5}$ volts to say 100 volts, and from values of resistance from $10^{-1}$ ohm to $10^8$ or $10^9$ ohms.

The range of validity of Ohm's law is an enormous help in circuit design. *Circuit models involving resistors, capacitors, inductors and operational amplifiers are transferrable.* They behave as real devices behave without changes in parameters. Largely for that reason, designs are inexpensive and robust.

Despite the simplicity of resistors, the actual current carrier in a carbon resistor, is unclear, at least to me. No one cares very much I suspect, because the device works nearly perfectly. The carrier of charge does not matter very much. What matters is the constitutive law that describes the relation of current, voltage and time. The constitutive law should satisfy conservation of mass, charge and current.

It is instructive to write the constitutive law for a resistor Ohm's law for only particle current, using a conservation of particle (mass) formulation, and then write it again for particle plus displacement current from one terminal to another. If you apply a step function of current (or potential for that matter), to the purely particle formulation, a paradox arises. The potential changes but the particle current flowing into the resistor from the left is exactly equal to the particle current flowing out on the right *at all times.* Why does the potential change if there is no accumulation of charge?

The paradox can be resolved in two (nearly) equivalent ways.

(1) The constitutive equation Ohm's law can be used with the extended definition of current that includes the displacement current. In that case there is continuity of generalized current, but there is *NOT* continuity of particle current at all times. The transient accumulation of particle current provides the charge that changes the potential. Paradox resolved.

(2) Alternatively, the circuit model of the resistor in Fig. 2 can be changed to have an explicit capacitor in parallel with it. In that case, the charge accumulates on the capacitor, and the resistor itself can have continuity of flux of particles at all times and follow Ohm's law using the current/flux of particles (and not the displacement current).

The charge accumulating on the capacitor creates and changes the electric field. Paradox resolved.

Maxwell himself repeatedly used capacitors in this spirit  to understand the role and significance of displacement current. Sections 102, 125, 199, and Chapters 8, are some examples, in [157]. We (following the insight and advice of Wolfgang Nonner) have used capacitors as a crude way to connect permanent charges (specifically, acid side chains in a channel protein away from the pore) and electrical potential in the pore of a channel protein, e.g., the potassium channel [61].

Current flow in wires can also be strange. The current carriers in a wire are delocalized electrons in the simple case of a single solid conductor of metal, and follow the simplest constitutive law of all at long(ish) times, say times longer than $10^{-5}$ sec. But most of our electronics occurs at times much shorter than that.

We confront the importance of the time variable. *The physical nature of current flow depends on time scales, even in wires* [115, 188]. The range of time scales in our technology is enormous, from more than one second to less than $10^{-9}$ sec. At the shorter times ($< 10^{-4}$ sec), the wires that must be used are often twisted pairs [13], each made itself of many very fine wires. Without twisting, these pairs of wires do not allow successful connections to the internet because rapidly changing signals are not carried reliably by single wires [201]. *At those short times, currents flow outside wires*, guided by the conductor, to be sure, but outside the conductor nonetheless [115, 201]. The twisting of wires is a necessity if they are to carry signals robustly and reliably so we can use them in our video devices and smartphones, even in old fashioned hard wired telephones [13] that only need to amplify audio signals heard by adult humans $< 10^{4}$ Hz.

Physical nature of current depends on time scale. The physical nature of current in almost any system depends on the time scale, and differs at different times as much as it differs in different devices. Constitutive equations depend on time. Different devices have different constitutive equations with different time dependence.

Again, charge movement is an abstraction, different at different times in one device, as well as different in different devices.

Charge Movement in Batteries. Batteries are present in Fig. 2 both as an isolated device and as the Ag‖AgCl⁻ interface between wires and Na⁺Cl⁻ solution in the conducting cylinder previously discussed.

I hesitate to describe a constitutive equation for the flow of charge in batteries in general because the flow is so very complex, different in different devices, and important for the practical daily use of batteries [36, 119] and its interaction with surface charges is also subtle and important [187]. It is enough to say here that current flow through electrochemical systems is carried by a wide range of charge carriers. Constitutive equations of different electrochemical systems differ and change nature dramatically with time and frequency and composition and concentration of ionic solutions, as well as electrical potential, and current flow. At short times, at $10^{-6}$ sec—that are still long and slow compared to the times important in computers—current from the Ag wire and AgCl electrode material into the Na⁺Cl⁻ solution is entirely displacement current lagging behind voltage but at the longer times characteristic of biological systems (greater than say 0.1 sec) the current is carried by a complex combination of Ag⁺ and Cl⁻ ions with negligible displacement current.

<u>We conclude that charge is indeed an abstraction</u> with different physical meanings in different systems and at different times. No one can visualize and no one knows—at least I do not know anyone who knows—why or even how this abstraction can be so perfectly conserved under all conditions and on all scales, from Angstroms to meters, from femtoseconds to seconds.

It is truly amazing to think of the changes in electric forces needed to accommodate and enforce continuity of current in salt water, vacuum capacitors, dielectric capacitors, vacuum tube diodes, semiconductor diodes, resistors, wires and batteries from atomic scale time $10^{-16}$ sec and distance $10^{-11}$ meters to the biological scale of seconds and meters from inside atoms (p. 8-9 of [59]) to intercontinental distances (in submarine cables) and interstellar space.

<u>General remarks about current flow</u>. The discussion of Fig. 2 leads to some general remarks about current flow. the interplay of conservation of matter and conservation of charge is different in each system, according to the constitutive law of the system. The idea of a current flow is an abstraction built to accommodate the physics of each system, while maintaining the main feature of electrodynamics, the exact conservation of current, at any one time, on any scale. Conservation of mass is certainly also followed in these systems, but in a more relaxed way. Significant deviations are allowed for a time. Mass can accumulate for a time without catastrophic results.

<u>Accumulation of mass does not have dramatic effects</u>**.** In the systems considered here, the accumulation of mass does not usually have dramatic effects. A one percent accumulation of mass has a small effect on chemical potentials, although of course exceptions can occur. There is no general law for the accumulation of matter. The effect depends on the constitutive law, and its interplay with conservation of matter and charge.

<u>Accumulation of charge has dramatic effects</u>. The accumulation of charge is different. A one percent accumulation of charge has a huge effect. As Feynman memorably mentions at the very beginning of his textbook [59], *one per cent excess of charge in a person at arm's length produces force enough to lift the earth!*

The accumulation of charge follows an exact law: the displacement current is the sum of all the other charge flows and it changes the rate of change of the electric field in a precise way so total current is conserved precisely. Continuity of current is obeyed and so such enormous forces do not occur. If continuity of current were not obeyed exactly, enormous forces would soon develop (see Appendix). Such forces are of course incompatible with life or laboratory experiments. Fortunately, accumulated charge is easier to deal with than accumulation mass: the vacuum displacement current does not depend on constitutive laws. It simply depends on the time derivative of potential and so can be calculated.

<u>Accumulated charge is much simpler than accumulated mass</u>. Accumulated charge has universal properties, *independent of the physical nature of the charge.* Particle and quasi-particle currents that accumulate at a junction change the time derivative of electric potential so the electrical potential carries away a displacement current. This displacement current is ***exactly equal to the sum of the other currents flowing into the junction*** without known error, to about one part in $10^{18}$. Continuity of current is exact *no matter what the constitutive laws* if current is re-defined to include displacement current. The displacement current (and equivalently $\partial V/\partial t$ 'take up the slack' so that Kirchhoff's current law (using the extended definition of current) is exact in one dimensional systems like a sequence of chemical reactions. No charge accumulates at all beyond that defined by the integral of the displacement current. The electrical forces and $\partial V/\partial t$ change so the displacement current exactly equals the sum of the other currents, and continuity of generalized current is exact.

<u>Accumulation of charge is special because it is universal</u>. The precise linkage between potential change, charge accumulation and displacement current is a special feature of electromagnetism because it is universal. It is a property of a vacuum—the constitutive equation of a vacuum, if poetic license is allowed. In this sense charge is more fundamental and more universal than mass, a fact which certainly came to me as a surprise.[3] But then I learned that charge is same at all velocities, 'Lorentz (relativistically) invariant', whereas mass is not. Mass depends on velocity and charge does not. Special relativity seems to make charge a more fundamental physical property than mass.

The reader may have difficulty visualizing the interactions that enforce conservation of the abstraction 'charge/current' in all these devices with all these properties over the entire time scale. I certainly do. *However,*

<u>Current flow (and charge) are conserved on and between all scales in all conditions</u>, even if we cannot visualize how that manages to be so. Experiments demonstrate that fact. Current does flow continuously without loss in a circuit. Consider a battery feeding a circuit. If a wire is cut far from the battery, current flow stops everywhere. The chemical reaction in the battery is disrupted on an atomic scale, by the (lack of) current flow meters away. Those of us living in colder climates have seen the effects when starting a car with another car's battery. We have learned to be careful because even a twelve volt battery can produce dangerous sparks in air, even though air has nearly infinite resistance ($> 10^{11}$ ohms for dry air), and is in that way, nearly a vacuum. Abstract charge is conserved 'exactly' even if we cannot visualize how that happens.

*Science often contains mysteries that cannot be visualized*—consider Maxwell's attempts to visualize his equations as properties of an ether. Science often poses questions that cannot be answered. Why is there no magnetic charge? Why is the charge on an electron $1.6 \times 10^{-19}$ coulombs? *Why is charge independent of velocity in special relativity when mass and distance and even time are not?* Why are physical laws invariant when locations $s$ move at constant velocity $\partial s/\partial t$ — special relativity—or at constant acceleration $\partial^2 s/\partial t^2$ —general relativity—but not when other time derivatives of location are constant, like a constant third derivative $\partial^3 s/\partial t^3$ or linear combinations of $\partial^n s/\partial t^n$, perhaps even fractional derivatives? As practical people, scientists cannot afford to just wait while we wonder about such things. Scientists wonder a bit *and* then move on, hoping our successors can do better than we have.

Biologists and engineers in particular cannot afford to linger on mysteries they do not understand. So many of those mysteries in biology have turned out to be caused by low resolution of our instruments, unable to resolve crucial structures. Think of Thomas Henry Huxley looking at the shortening of the striations of muscle [105] that were not understood until one of his grandsons (Andrew Huxley), studied them many years later working in Cambridge UK [102]. Think of Lee DeForest using vacuum tubes without understanding how they work.

Biologists and engineers cannot afford to wait to understand everything. They must isolate the mysteries and move on to study other things. Here, we move on to discuss devices and the theories and simulations used to understand them.

---

[3] No universal precise linkage exists between mass accumulation, rate of change of chemical potential, and flux, for example. The linkage for mass depends on the details of the constitutive equation, and *on all components of an electrolyte*, if the system is a nonideal ionic solution, like those that life requires.

<u>Biological Implications of Continuity of Current</u>. It is important to note that the continuity of current law has important biological implications in systems more general than a series of chemical reactions. Continuity of current law implies the cable equations (called the telegrapher's equation in the mathematics literature), see derivation from the three dimensional theory in [5, 55, 170, 172] and p. 218-238 of [118]. The cable equation [94] is the foundation of the Hodgkin Huxley model [93, 96, 101, 103, 104] of the action potential of nerve and muscle fibers. Kirchhoff's current law links the atomic properties of ions, the molecular properties of ion channels, and the centimeter scale spread of current and potential that creates the propagating action potential in nerve fibers meters in length.

In short or round(ish) cells, or in organelles like mitochondria, continuity of current forces coupling between multiple pathways of current crossing membranes, *even if the currents are carried by different ions, or by electrons, through different structures* nanometers apart in the membrane of the finite cells or organelles.

The flux coupling characteristic of active transport systems—including the coupled flows in chemi-osmotic systems that perform oxidative phosphorylation or photosynthesis—might arise in this manner. *Coupling of flows of charges, whether electrons or ions, is an unavoidable consequence of GLOBAL conservation of charge and current*, of Kirchhoff's current law GLOBALLY enforced in three dimensions, and not a consequence of local chemical interactions, just as coupling of membrane currents with axial currents in a nerve fiber is an unavoidable consequence of Kirchhoff's current law, not of local chemical reactions.

It is interesting to compare the incorrect chemical theory of nerve propagation of Nobel Laureate A.V. Hill [86] with the correct electrical theory of the then undergraduate [91, 92], later Nobel Laureate [93, 96, 101, 103, 104], Alan Hodgkin. Kirchhoff's current law in the form of the cable equation [94] was the key to Hodgkin's understanding. The classical voltage clamp experiments were designed to remove difficult terms, and isolate membrane terms, in the cable equation—personal communication, A.L. Hodgkin, 1961—that today we know describe ion channels, opening, closing and conducting [89, 165, 186].

<u>Cable equation links movement of atoms inside channel proteins to macroscopic current</u> flow that produces nerve propagation of the macroscopic electrical potential, the nerve signal that spreads meters. Macroscopic potentials modify atomic movements involved in gating and conduction. Atomic movements create the macroscopic electrical potentials.

Equations of the electric field are true on all scales and so allow a unique linkage between models of atomic motion, protein behavior, and macroscopic propagation of electrical signals. I suspect linkage equations of this type—valid on all scales—will be needed to make any multiscale analysis robust and transferrable, if it reaches from atoms to meters, from femtoseconds to minutes as models of nanodevices must.

<u>Models, Devices, Effective Parameters, and Transferrable Theories</u>. Parameters of models or devices can often be chosen so an incomplete theory or simulation describes a system in one set of conditions but not another. Experiments often show that rate constants must be adjusted dramatically as conditions change, and the adjustments can rarely be predicted ahead of time by theory.

Chemistry and biology are filled with examples of *non*-transferable models. Chemical reactions follow rate equations, but the rates are not constant, not independent of one another, as conditions change, even though theory assumes they should be [47]. Biology describes enzymes with one set of parameters but finds those are changed when conditions change and attributes that somewhat mysteriously to 'allosteric effects' and conformation changes.

<u>Non-transferrable theories have limited use</u>. Biology and much of chemistry works under a wide range of conditions and so incomplete theories with effective parameters have limited use. Even if sensible, even if valid, theories (and simulations) with effective parameters like these are not accurate enough to design robust devices. By leaving out something important, those theories or simulations leave out an energy term that is almost certain to vary with conditions. The resulting effective parameters change in large and unpredictable ways.

Incomplete theories and simulations are not very useful over a range of experiments and conditions. Incomplete theories are not likely to be transferable (from one set of conditions to another) in the language of the chemistry literature. Devices designed from incomplete theories or simulations are unlikely to be robust or work well under a range of conditions. Biological systems analyzed with non-transferable theories (or simulations) are unlikely to be realistic in general because biological systems almost always work in a range of ionic concentrations different from those used in the laboratory.

<u>Simulations must deal with trace $Ca^{2+}$</u>. Biological systems usually work in mixtures with a range of $Ca^{2+}$ concentrations, in which $Ca^{2+}$ concentration has important practical effects, often turning systems on or off or controlling their rate monotonically. Simulations and theories in biology have limited use until they are calibrated so we can be sure they actually are correct in the range of conditions and $Ca^{2+}$ concentrations the biological system uses. Simulating $Ca^{2+}$ activity in pure solutions is a challenging problem [189]. Simulating $Ca^{2+}$ activity in biological mixtures in the $10^{-8}$ to $2 \times 10^{1}$ molar concentrations that are physiological, has not been attempted as far as I know. ($10^{-8}$ concentrations of $Ca^{2+}$ are found inside most cells. $2 \times 10^{1}$ concentrations of $Ca^{2+}$ are found in and near ion channels, nucleic acids, and enzyme active sites, where the chemistry of life is catalyzed and controlled.

Biological and chemical science will benefit enormously if theories and simulations can be made transferable, using one set of parameters to describe systems in a range of conditions, as many physical and most engineering theories and simulations do. I believe the law of mass action must be extended to conserve current before theories and simulations can be made transferable from condition to condition, from physics, to chemistry to biology, using only mathematics.

<u>How to extend the law of mass action so it conserves current</u>? An obvious way to extend the law of mass action is to include activities and electrical potential in the rate constant to 'right the rates' by making them (typically exponential) functions of potential [4, 7, 78, 79, 122]. This in fact has been done for a very long time in the study of reactions at the electrodes of electrochemical cells and recently in other ways in the treatment of the formation of concrete [98, 112, 148]. The Butler-Volmer and Tafel equations [7] include electrical potential in rate constants in an empirical way with limited [78, 79] but real success.

Success is limited I suspect because difficulties of embedding an electrical potential in rate constants are formidable if current flows. We must 'fix the fields' so they are global and depend on current flow everywhere. Otherwise, they cannot conserve charge flow and support continuity of current as required by Maxwell's equations.

*The way to 'right the rates' is to fix the fields,*
*everywhere.*

**Additional Perspectives**. Some additional general remarks may be helpful.

(1) A thermodynamic treatment is clearly impossible since the goal is to calculate large fluxes and currents that do not occur in a thermodynamic system at equilibrium by definition without flows.

(2) A rate treatment of the frictional treatment of flux over a large potential barrier is needed. The classical Brownian motion problem of Kramers [60, 81] is a necessary step forward, even if it is inconsistent because it does not compute the potential barrier from the charges in the system. Kramers' treatment need not be restricted to large barriers. A simple expression for rates over any shape barrier is available [40] and needed [6] because so few barriers are both symmetrical and large as required in classical high barrier approximations.

(3) The rate constants over one barrier must depend on the electrical potential in far locations. Otherwise, interrupting current flow in a far location cannot interrupt current locally. This requirement implies that the electrical potential must be determined by a global equation like Poisson's equation, including boundary conditions far from the individual chemical reactions and barriers. The barriers of Kramers' model are variable and not constant as conditions change, including conditions far from the barrier itself. The fields and energy landscapes change and it is indeed their change that allows continuity of current (1) to be satisfied.

(4) The rate constants over one barrier are likely to depend on concentrations in other places because the solutions containing the reactants are not ideal[3,4]. A general characteristic of nonideal solutions is that 'everything depends on everything else'. More specifically, the activity of one reactant (the free energy per mole) depends on the concentrations of other species in practice, as well as in principle.

(5) A general theory of all nonequilibrium processes is not likely to be useful: a general theory has to describe too much. A general theory must include hydrodynamic behavior of considerable complexity, since aqueous solutions are fluids satisfying the Navier Stokes equations of fluid mechanics. A general theory would also include explosions since they occur with regrettable frequency at electrodes of electrochemical cells, when $H_2$ gas is generated (inadvertently) by an overvoltage.

Fortunately, nonequilibrium processes in biology and much of technology occur in ionic solutions in which atomic motion is heavily damped. That damping ensures that the distribution of velocities is a displaced Maxwellian [56], as it is in semiconductor devices [15, 84, 210]. The displaced Maxwellian has non-zero mean velocity and so allows flux and current through the system from power supplies to outputs. That flux and current is enough to produce the very nonlinear devices of our digital (semiconductor) technology [109, 152, 156, 193, 210], and nonlinear phenomena like the propagating signal of the nervous system, the action potential [101, 103, 104].

**Boundary conditions are needed to extend the law of mass action** to deal with the outside world. Equilibrium statistical mechanics and thermodynamics were designed to avoid the complexities of boundary conditions using the 'thermodynamic limit' to allow analysis. But when current flows, interactions with the outside world are unavoidable, the thermodynamic limit is not appropriate, and boundary conditions describing those interactions are needed.

It is difficult if not impossible to deal in general with boundary conditions for the law of mass action, and chemical reactions, because of the wide variety of chemical interactions and physical geometries captured in a phase space of very high dimension. Boundary conditions are easier to deal with in three dimensional physical space, where the law of mass action is widely, nearly universally used to describe ion motion through narrow channels [40, 87-89].

Current in channels. Current flow through a 'hole in a (insulating) wall' is a subject of great significance because the hole in the wall allows control of the current. Holes in proteins, holes in membranes, and channels in field effect transistors are all nanovalves providing essential functions to a large fraction of biology and semiconductor technology, and of great interest in electrochemistry. Indeed, biological channels are nearly picovalves allowing a handful of atoms to control macroscopic flow. It is hard to imagine something smaller. We consider models of current flow through such holes and show how boundary conditions can be applied to nanovalves and systems of this type.

We do not consider simulations here because they do not deal with the essential features of these systems. Simulations have considerable difficulties in computing the actual currents through such systems as they are controlled by handfuls of atoms. The currents are macroscopic phenomena, occurring in the world of milliseconds to minutes, and spatial dimensions from say $10^{-9}$ to 1 meter. The currents are controlled by atomic scale structures in biology and near atomic scale ($10^{-8}$ meter) structures in semiconductors and electrochemistry. Changing a few atoms changes macroscopic currents as is shown in biophysical experiments (usually involving site directed mutagenesis) every day. The macroscopic currents are driven by chemical and electrochemical potentials involving large numbers ($>10^{15}$) atoms in most cases. Simulations must then compute macroscopic scale inputs and outputs while preserving atomic scale spatial resolution of the controlling atoms. These issues must be all dealt with at once, because they all exist at once in the systems of interest, and they must be dealt with accurately, because valves typically depend on the balance of nearly equal forces, e.g., electrostatic and diffusion. Each force must be quite accurately calculated because it is the difference that controls function. It will be some time before simulations can surmount these problems and be used to make practical devices [38, 155, 176]. Meanwhile, we use the mesoscopic approach which has been so fruitful in computational electronics [152, 195, 210] where atomic scale simulations are rarely if ever used. The key to the mesoscopic approach is the choice and treatment of correlations. Not all correlations can be handled (see the infinite series in [192] which of course has not been shown to converge).

We consider the Poisson equation and use the treatment in Barton [11] (p.168) to illustrate the issues in connecting a nanovalve to the outside three dimensional world.

$$\nabla^2 \phi(\mathbf{r}) = -\rho(\mathbf{r}) \tag{12}$$

An integral ('Kirchhoff') representation of equation inside the nanovalve is $K_{vol}(\mathbf{r}) + K_{surf}(\mathbf{r})$, where $F_{vol}(\mathbf{r})$ involves the usual free space Green's function $G_{free}(\hat{\mathbf{r}} \mid \mathbf{r}) = 1/4\pi(\hat{\mathbf{r}} - \mathbf{r})$

$$K_{vol}(\mathbf{r}) = \int_{vol} d\hat{V}_{vol} \rho(\hat{\mathbf{r}}) G_{free}(\hat{\mathbf{r}} \mid \mathbf{r}) \text{ where } G_{free}(\hat{\mathbf{r}} \mid \mathbf{r}) = \frac{1}{4\pi} \frac{1}{\hat{\mathbf{r}} - \mathbf{r}} \tag{13}$$

$$K_{surf}(\mathbf{r}) = \int_{surf} d\hat{S}_{surf} \, \partial_n \phi(\hat{\mathbf{r}}) \hat{G}_{free}(\hat{\mathbf{r}} \mid \mathbf{r}) - \int_{surf} d\hat{S}_{surf} \phi(\hat{\mathbf{r}}) \partial_n \hat{G}_{free}(\hat{\mathbf{r}} \mid \mathbf{r}) \qquad (14)$$

Here we use Barton's notation for the normal derivative $\partial_n \phi(\hat{\mathbf{r}})$ of the potential (for example) as a function of the location of the source $\hat{\mathbf{r}}$. Inside the nanovalve the volume $K_{vol}(\mathbf{r})$ and surface terms $K_{surf}(\mathbf{r})$ add to give the solution $\phi(\mathbf{r})$ of Poisson's equation (12)

$$
\left.
\begin{array}{ll}
K_{vol}(\mathbf{r}) + K_{surf}(\mathbf{r}) = \phi(\mathbf{r}) & \mathbf{r} \text{ is inside the nanovalve} \\[1em]
K_{vol}(\mathbf{r}) + K_{surf}(\mathbf{r}) = 0 & \mathbf{r} \text{ is outside the nanovalve}
\end{array}
\right\} \qquad (15)
$$

The free space Green's function $G_{free}(\hat{\mathbf{r}} \mid \mathbf{r})$ is of course the average of the free space potential in atomic scale simulations.

The special properties of nanovalve depend a great deal of the properties of its surface because nanovalves are so small. Structural biology determines the surface of the protein nanovalve and the amino acids forming that surface. Physics determines the surface Green's function $\partial_n \hat{G}_{free}(\hat{\mathbf{r}} \mid \mathbf{r})$ and thus the surface normal derivative $\partial_n \phi(\hat{\mathbf{r}} \mid \mathbf{r})$ and the surface charge.

Determining these properties is the goal of analysis of specific nanovalves and does not concern us here. It is enough to mention that these properties can be determined from experiments analyzed by the mathematical solution [22] of the appropriate inverse problem. The spatial distribution of structure and permanent structure can be determined from measurements of current voltage relations in a wide range of ionic conditions, concentrations, and voltages $\pm 6 \, k_B T/e \, (150 \, \text{mV})$. The large amount of accurate data allows accurate solution of the inverse problem.

<u>Connection of the nanovalve to the external world is what concerns us here</u>. In mathematical language, the problem reduces to the Kirchhoff representation of the end of the channel, written from eq. (14) by isolating the surface of the ends from the rest of the structure.

$$K_{end}(\mathbf{r}) = \int_{end} d\hat{S}_{end} \, \partial_n \phi(\hat{\mathbf{r}}) \hat{G}_{free}(\hat{\mathbf{r}} \mid \mathbf{r}) - \int_{end} d\hat{S}_{end} \phi(\hat{\mathbf{r}}) \partial_n \hat{G}_{free}(\hat{\mathbf{r}} \mid \mathbf{r}) \qquad (16)$$

In general this problem can be complex involving interactions of all sorts between the interior of the nanovalve or channel and the external world. Indeed, in some biological channels (calcium channels) this may be important (although so far little studied). In general, however, nanovalves are devices designed to work reasonably robustly and independently of the world around them. Robust devices need to be transferable from one place to another and so the complex interactions are minimized by the design and evolution of the systems. If we oversimplify to make the point clearly: nanovalves are exceedingly narrow where they allow control but widen dramatically outside that region so 'resistance' to flow is concentrated in the narrow region. Control is robust, available in a wide range of surrounding conditions.

In the nanovalves of semiconductor technology, buffer regions isolate the nanovalve and allow it to have robust properties. The connection to the outside world is through a buffer region of semiconductor separating the metallic contact from the nanovalve itself. In the nanovalves of biology, the buffer regions are the antechambers of the channel and the surrounding ionic baths between the Ag | AgCl wire (or salt bridge) and the ion channel. These regions are designed (or evolved) so the current in the buffer region is 'Ohmic' independent of time during the function of the nanovalve. These regions are designed to minimize the layers of charge of eq. (16) that create undesirable complex behavior not easily controlled by the valve itself.

The connection of the nanosystem to the outside world is the current flowing in and out of the channel, and of the potential at the ends of the channel. The current flowing in and out of the channel are not equal because of transient charge storage ('capacitive' properties, famously voltage dependent and nonlinear) are large and significant in semiconductor valves [76, 108, 173, 203, 210]. In biological channels, voltage dependent charge storage phenomena are present as well, where they are called 'gating currents' [3, 14, 191]. These nonlinear displacement currents flowing in channels and associated structures are small but they are controllers of biological function of great importance, and link the motions of handfuls of atoms to macroscopic function. These currents are small because they are produced by motions of a small number of charges compared to the total number of charges in the system. They can be measured because continuity of current guarantees that charge movements arising in conformation changes must also flow in the electrodes and circuits connected to them [49].

The currents flowing in and out of the nanovalves are the connection to the outside world and need to be included in the analysis of the nanovalve itself. These are the currents that are continuous. The currents are the same everywhere in a series connected system like those described in eq. (6).

The rate constants of mass action models of particle movement can be connected to the external world using a theory that accounts for current flow everywhere, along with diffusion, and perhaps migration as well.

<u>Rate Constants for Nanovalves</u>. Analytical expressions for the rate constants for movement of charged particles in channel structures can be derived in quite a general way, starting with Langevin equations for the thermal motion of ions [40, 56, 164, 199, 200]. It is necessary to use the full Langevin equation (including second derivatives with respect to location) if the treatment is to allow two boundary conditions (i.e., electrochemical potential on the inside and also the outside of the channel) and allow macroscopic flux. (If one uses only the Smoluchowski, high friction version, of the Langevin equation, with only first order spatial derivatives, the distribution of velocities of particles has mean zero and no net flux.) Trajectories can be doubly conditioned, allowing separate boundary conditions for the two sides of the channel, and the resulting multiple integrals can be performed analytically, somewhat surprisingly, to give the expressions simulated in ref [6], derived in ref [56], as shown in ref [40]

$$L_{eft} \; \underset{k_b}{\overset{k_f}{\rightleftharpoons}} \; R_{ight} \tag{17}$$

where

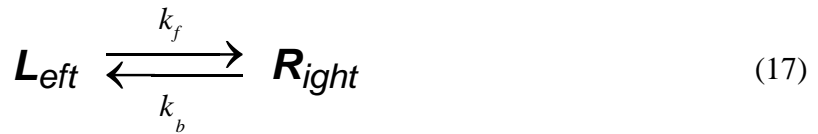

$$J_k = \overbrace{l \cdot k_f \cdot C_k \left( L_{eft} \right)}^{\substack{\textit{Unidirectional Efflux} \\ J_{out}}} - \overbrace{l \cdot k_b \cdot C_k \left( R_{ight} \right)}^{\substack{\textit{Unidirectional Infflux} \\ J_{in}}}. \tag{18}$$

The rate constants are conditional probabilities derived using the theory of stochastic processes from the properties of doubly conditioned Brownian trajectories.[56, 164, 199, 200]

$$J_k = \overbrace{\underbrace{C_k(L)}_{\substack{\text{Source}\\\text{Concentration}}} \left(\underbrace{\frac{D_k}{l}}_{\substack{\text{Diffusion}\\\text{Velocity}}}\right) \underbrace{\text{Prob}\{R|L\}}_{\substack{\text{Conditional}\\\text{Probability}}}}^{\text{Unidirectional Efflux}} - \overbrace{C_k(R)\left(\underbrace{\frac{D_k}{l}}_{\substack{\text{Channel}\\\text{Length}}}\right)\text{Prob}\{L|R\}}^{\text{Unidirectional Influx}} \tag{19}$$

or

$$J_k = \overbrace{\overbrace{l \cdot k_f C_k(L_{eft})}^{J_{out}}}^{\text{Unidirectional Efflux}} - \overbrace{\overbrace{l \cdot k_b C_k(R_{ight})}^{J_{in}}}^{\text{Unidirectional Influx}}, \tag{20}$$

where

$$\left.\begin{array}{l} k_f \equiv \dfrac{J_{out}}{C_k(L_{eft})} = k\{R_{ight}|L_{eft}\} = \dfrac{D_k}{l^2}\text{Prob}\{R_{ight}|L_{eft}\} = \dfrac{D_k}{l^2}\dfrac{\exp(z_k F V_{trans}\,/\,RT)}{\frac{1}{l}\int_0^l \exp(z_k F \phi(\zeta)\,/\,RT)d\zeta}; \\[3em] k_b \equiv \dfrac{J_{in}}{C_k(R_{ight})} = k\{L_{eft}|R_{ight}\} = \dfrac{D_k}{l^2}\text{Prob}\{L_{eft}|R_{ight}\}\dfrac{D_k}{l^2}\dfrac{1}{\frac{1}{l}\int_0^l \exp(z_k F \phi(\zeta)\,/\,RT)d\zeta}. \end{array}\right\} \tag{21}$$

$R$ is the gas constant, $F$ is Faraday's constant, $T$ is the absolute temperature, $V_{trans}$ is the electrical potential across the channel not including potential drops outside the channel.

The coupling to the long range fields and flows is through the expressions for the electrical potential $\phi(x)$ because the regions outside the channel are decently ohmic [2, 80, 149, 150, 171]. First order dependence on concentration in the bath can be described by changes in the concentrations on left $C_k(L)$ and right $C_k(R)$. The changes in concentration can often be described adequately this way because they are slow and small. The potential is computed in all space (channel and surrounding baths) by a consistent theory, in one flavor or another. **PNP** [12, 24, 41, 52, 193] deals with the motion of point charged particles; **EnVarA [50]** deals with spherical particles, that can diffuse, migrate, or flow by convection, **steric**- **PNP** [97] is an approximate version much easier to compute, and **PNP - Fermi** [145-147] deals with finite size by enforcing a Fermi distribution that prevents over filling and accounts for saturation of space (by spheres), and no doubt there are many other appropriate models in the vast literature (which includes semiconductor applications, ionic solutions, ion channels, and formation of concrete). These models produce potential profiles that automatically change with conditions so rate constants change and produce currents that are continuous and satisfy Maxwell equations.

Nanovalves are obviously a small subset of the applications of the law of mass action, but it is some comfort to see how consistent analysis can be done in this case so the law of mass action and continuity of current can both be satisfied. The general case is much harder. It is difficult to grasp all the dimensions of chemical reactions at this stage. One can reach in that direction by studying specific chemical reactions, where simple representations (involving one dimensional reaction paths) are enough to describe important phenomena.

<u>Variational methods can extend the law of mass action</u>. If the goal is to build transferable systems, so we can build robust devices, as in electronic technology, we must use a mathematics that allows interactions of charges and fields, currents and fluxes and flows of solvent, extending from atomic to macroscopic scales.

Variational methods are designed to deal with systems with multiple forces and flows, in which interactions are unavoidable and complex. In these systems, interactions must be included in all analysis. Otherwise, theories have more adjustable parameters than can be determined experimentally and still cannot deal with a range of conditions, because interactions change with conditions in ways too complex for ordinary theories.

<u>Theories of ionic solutions have difficulties</u>. Sadly, theories of ionic solutions seem to have these difficulties Theories of ionic solutions need a large number of adjustable parameters and still cannot describe biological solutions (for example).

The first step in analysis seems to be the identification of properties of single ions from measurements of solutions—that always contain at least two types of ions, cation and anion, because of electrical neutrality. The identification of properties of single ions remains 'elusive' even in Hünenberger and Reif's [100] six hundred pages and more than two thousand references. Hünenberger and Reif's title itself characterizes single ion solvation *as elusive even in the infinitely dilute solutions they consider*. The ionic solutions needed to sustain life (and used in much of electrochemical technology) are much more concentrated and have many more and stronger interactions among solutes; interactions between solutes and water; and interactions with far field boundary conditions. They do not resemble infinitely dilute solutions [39-45].

References [39, 44, 63, 64, 121, 123, 124, 127, 174, 194, 218] draw particular attention to the difficulties and remind readers that almost all biology and electrochemistry occurs in solutions more concentrated than 0.1M, often in solutions much more concentrated. Solutions tend to be most concentrated where they are most important, near electrodes (in electrochemistry), in and near nucleic acids, binding sites of proteins, enzyme active sites [114], ion transporters and ion channels [17, 18, 72-74]. Almost all of biology occurs in ionic mixtures and involves flow and so are described particularly poorly by existing theories and simulations [39, 41, 42, 44-46].

The sad limitations of our understanding of ionic mixtures, like those in which all of biology occurs, are not widely known, and so embarrassing that many do not want to know of it, as I did not for many years. It is necessary then to document the frustration by quotations from leading workers in that field.

The classical text of Robinson and Stokes [180] is still in print and widely used. It is a book not noted for emotion that still gives a glimpse of its authors' feelings of frustration (p. 302)

"In regard to concentrated solutions, many workers adopt a counsel of despair, confining their interest to concentrations below about 0.02 M, ... "[4]

---

[4] Note that almost all biology and electrochemistry occurs in much more concentrated solutions.

In a recent comprehensive treatment [124] of nonideal properties of solutions, the editor Werner Kunz says (p. 11 of [125])

> "It is still a fact that over the last decades, it was easier to fly to the moon than to describe the free energy of even the simplest salt solutions beyond a concentration of 0.1 M or so."[4]

New mathematical tools are needed to resolve a stalemate existing since the 1920's. The powerful tools of variational calculus automatically deal with interactions that vary dramatically with conditions. If the mathematics does not deal with interactions, those interactions will not be computed correctly and will wreak havoc with theories that are based on algebraic descriptions of interactions or theories based on the law of mass action with constant rate constants.

Energy Variational Approach. The energy variational approach *EnVarA* is *defined* by the Euler-Lagrange process [57] as generalized by Liu, and colleagues, into an energy/dissipation functional. The generalized functional combines two variations yielding a single set of Euler Lagrange equations (in Eulerian coordinates of the laboratory) using push back and pull forward changes of variables [50, 62, 184, 213, 215]. In this way, *EnVarA* can deal with dissipation (friction) and ionic solutions.

*EnVarA* describes conserved energy using the classical Hamiltonian variational principle of least action described in textbooks of mechanics. It deals with friction using the Rayleigh dissipation principle described in textbooks of irreversible thermodynamics. When combined, *these principles allow energy to be degraded into entropy as matter and charge flow in frictional materials* [216], like electrolyte solutions.

Ionic solutions experience friction because an ionic solution is a condensed phase essentially without empty space. Ice floats on water, so liquid $H_2O$ has greater density than solid $H_2O$ and presumably less space between its atoms than the solid. Atoms and molecules in a condensed phase cannot move without colliding. Collisions randomize originally correlated motions and make them into the kind of randomized motion that we call heat [19]. The variance of the displacement is what we call temperature. The macroscopic names for the conversion of translational to randomized zero-mean (nearly) Brownian motion, are dissipation and friction.

Energetic variational methods are particularly useful because they allow multiscale derivation of partial differential equations (and far field boundary conditions) from physical principles when multiple fields are involved, like convection, diffusion, steric exclusion, and migration in an electric field. Energetic variational principles have recently become available for systems involving friction [62, 106, 215], that is to say, for systems involving ionic solutions [50, 99, 161, 213].

Energetic variational principles combine the full power of the Navier Stokes equations (a) with either a Lennard Jones representation of finite size ions or (b) with a density functional theory of ionic solutions built from Rosenfeld's density functional theory of liquids [50, 107].

Computations must be done in three dimensions because spheres do not exist in one and two dimensions. These theories and their simplifications [97, 128, 142] are difficult to compute in three dimensions because of the steeply singular forces used to ensure that atoms do not overlap. Overlap must not be permitted because spheres cannot overfill space: space can be saturated with spheres. Saturation effects are a main cause of nonideality, particularly in the extremely crowded conditions in and near enzyme active sites, ionic channels, nucleic acids, and the working electrodes of electrochemical cells where twenty molar solutions are not uncommon [114]. (As a rule of thumb, ions are crowded and electric fields are largest [66, 85, 202] where they are most important in technology and biology.)

If saturation is described by a Fermi-like distribution—as recently derived for spheres of unequal sizes [143, 144]—some of these difficulties can be attacked [145-147]. A fourth order partial differential equation can be written [145] which is easily integrated in three dimensions, after it is reduced to a pair of second order partial differential equations (with carefully defined boundary conditions) and computed with appropriate numerical methods.

But it is still not clear how best to apply any of these methods to chemical reactions (Fig. 1) described by the law of mass action with rate constants extended to be functions or functionals and not constants.

<u>And that is not a bad place to move this essay towards its coda</u>.
We now have the tools, and we now see the goal—a global treatment combining conservation of mass in chemical reactions described by the law of mass action with conservation of charge flow and current described by Kirchhoff's current law everywhere.

***Now we have to do the work.***
***We have to actually implement consistent models and see how well they do.***

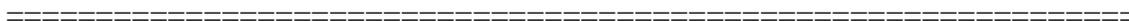

# <u>Coda</u>
*Our grasp must be sure, but our reach should exceed our grasp,*
*as we do our science.*

We must grasp both charge/current conservation and mass action before we can produce robust theories (or simulations) of chemical reactions in ionic solutions that successfully use one set of parameters in a range of conditions, and include the global properties of the electric field.

Correct calculations are needed because there is no engineering without numbers and accurate computations [38, 155, 176]. Calculations from theories and simulations—of electronic or ionic or biologically inspired devices—must be checked and calibrated against known results. *Otherwise devices built from those calculations will not work* [39, 41, 42, 44-46]. If theories and simulations of electrical devices were not robust, if parameters had to be changed as conditions changed, our electronic technology would be severely limited, to say the least.

Chemical reactions involving current flow must be within our theoretical grasp before we can develop transferable theories. Then can we build devices that perform as expected, as electronic devices usually do. Only then can we expect exponential growth in molecular engineering whether biological or technological. It seems no coincidence that exponential growth in electronic technology came after scientists had a secure grasp of global electrostatics and the *PNP* equations of electrodiffusion in semiconductors.

Let us hope that energetic variational methods can grasp ions and chemical reactions in water as well and as successfully as *PNP* has grasped the useful properties of holes and (semi)-electrons in silicon and germanium.

# Acknowledgement

Fred Cohen, Eduardo Rios, and Brian Salzberg provided most useful criticisms and suggestions of an earlier version of this paper. Many thanks! Ardyth Eisenberg edited the manuscript with vigor and love, for which I am most grateful, as it motivates me every day.

**Appendix**
# When does mass action conserve current?
## What are the effects of discontinuity in current flow?

<u>Size of discontinuity of current flow</u>.

The difference in current in two sequential chemical reactions is shown in eq. (22). The difference is the discontinuity of current, the violation of Kirchhoff's law of continuity of current flow. The difference can be zero only for special circumstances. The difference is not zero in general circumstances, nor robustly.

$$\frac{I_{XY} - I_{YZ}}{F} = z_X \cdot k_{xy}\left[X\right] - z_Y \cdot k_{yx}\left[Y\right] - z_Y \cdot k_{yz}\left[Y\right] + z_Z \cdot k_{zy}\left[Z\right] \qquad (22)$$

<u>When are both conservation laws are satisfied</u>? When $I_{XY} - I_{YZ} = 0$ in eq.(22) –eq. (25), the law of mass action is consistent with Kirchhoff's current law and conservation of mass and conservation of charge/current are all satisfied.

<u>Special cases</u>. Units of current $I_{XY} - I_{YZ}$ here are $\left(\text{cou/sec}\right)/\text{liter}$.

<u>Special Case A</u>: If all *concentrations are set equal to one*, the currents (in the special case with a tilde)

$$\frac{\tilde{I}_{XY} - \tilde{I}_{YZ}}{F \cdot 1\frac{\text{mole}}{\text{liter}}} = z_X \cdot k_{xy} - z_Y \cdot k_{yx} - z_Y \cdot k_{yz} + z_Z \cdot k_{zy}; \quad \text{concentrations} = 1\frac{\text{mole}}{\text{liter}} \qquad (23)$$

<u>Special Case A\*</u>: If we also set all *charges equal to one*, along with *concentrations equal to one*,

$$\frac{\hat{I}_{XY} - \hat{I}_{YZ}}{F \cdot 1\frac{\text{mole}}{\text{liter}}} = k_{xy} - k_{yx} - k_{yz} + k_{zy}; \text{concentrations} = 1\frac{\text{mole}}{\text{liter}}; z_X = z_y = z_Z = 1 \qquad (24)$$

In this special case of eq. (24), labelled *A\**, asymmetry (net difference) in rate constants determines the discontinuity of current, the violation of Kirchhoff's current law.

<u>Special Case B</u>: Alternatively, we can set all *rate constants* and all *concentrations equal to one*,

$$\frac{\hat{I}_{XY} - \hat{I}_{YZ}}{F \cdot 1\frac{\text{mole}}{\text{liter}}\frac{1}{\text{sec}}} = z_X - z_Y - z_Y + z_Z; \text{concentrations} = 1\frac{\text{mole}}{\text{liter}}; \text{rate constants} = 1\frac{1}{\text{sec}} \qquad (25)$$

In this special case of eq. (25), labelled ***B***, asymmetry (net difference) of charges (valences) determines the discontinuity of current, the violation of Kirchhoff's current law.

<u>Asymmetry of parameters violates conservation of charge/current</u> and produces discontinuity in current from device to device, i.e., it produces accumulation of charge, with sizable effects, as shown next and are to be expected, given the strength of the electric field as discussed in the text.

<u>Size of effects</u>. To estimate the effect on electrical potential *V,* we need to know the size of the system. Imagine a spherical capacitor of radius *R*. Its capacitance to ground—or coefficient of the self-energy, if one prefers the phrase in the chemical literature—is $C = Q/V = 4\pi\varepsilon_r\varepsilon_0 R$ or numerically $4\pi \times 8.85 \times 10^{-12}$ [farad/meter]$\varepsilon_r R = 1.11 \times 10^{-10} \varepsilon_r R$ [farad] where $\varepsilon_r$ is the relative dielectric coefficient, about 80 in water solutions at longish times (say $> 10^{-5}$ sec). Then, a 1 nm radius capacitor with dielectric coefficient 80 has capacitance of $8.9 \times 10^{-18}$ farads.

Small charges produce large voltages in such a tiny capacitor. Even the charge on just one ion ($1.6 \times 10^{-19}$ cou) would produce 18 mV, large enough (compared to the thermal potential of 25 mV) to have a noticeable (~50%) effect in theories and simulations, because $\exp(-18/25) = 0.49$. (Components of rates often vary exponentially according to $\exp(-V / k_B T)$.) A unit discontinuity in current $I_{XY} - I_{YZ}$ in eq. (23) − (25) lasting for a second would produce a voltage of $V = Q/C = (1/F)/(8.9 \times 10^{-18}) = 1.14 \times 10^{12}$ volts.

Of course, 1 second is a long time for current to flow. If current flowed on a biological time scale, for 1 msec in a structure 1 nm in radius, with dielectric coefficient 80, the electrical potential would be much less, 'only' $10^9/\varepsilon_r$ volts, somewhat less than $10^7$ volts at low frequencies in water. Current flow of even a microsecond, would produce nearly ten thousand volts.

<u>Rate of change of potential</u>. We can also look at the effect on the rate of change of potential. The discontinuity of current is connected to the rate of change of potential by a version of Coulomb's law

$$\frac{\partial V}{\partial t} = \frac{1}{C} I \qquad (26)$$

If we apply this formula to the discontinuity of current in the special case of eq. (24), labelled *A\**, we can estimate how quickly that discontinuity of current would change the potential

$$\frac{\partial V}{\partial t} = \frac{\hat{I}_{XY} - \hat{I}_{YZ}}{F \cdot 1_{\text{(mole/liter)}}} \cdot \frac{F}{C} = \frac{\hat{I}_{XY} - \hat{I}_{YZ}}{F \cdot 1_{\text{(mole/liter)}}} \cdot \frac{F}{4\pi\varepsilon_r\varepsilon_0 R} = F \frac{k_{xy} - k_{yx} - k_{yz} + k_{zy}}{4\pi\varepsilon_r\varepsilon_0 R} \qquad (27)$$

For a capacitor $R = 1$ nm with $\varepsilon_r = 80$ and capacitance of $8.9 \times 10^{-18}$ farads (see above)

$$\frac{\partial V}{\partial t} = 1.1 \times 10^{17} \times F\left(k_{xy} - k_{yx} - k_{yz} + k_{zy}\right) \quad \text{in volts/sec} \qquad (28)$$

In other words, the breakdown voltage (~ 0.2 volts) of membranes and proteins would be reached in $1.1 \times 10^{-22} / (k_{xy} - k_{yx} - k_{yz} + k_{zy})$ sec. The breakdown voltage for matter in general (say $10^6$ volts) would be reached very quickly.

We conclude that:

| |
|---|
| Failure of the law of mass action to conserve current is likely to have noticeable effects. |